\documentclass[a4paper,12pt]{article}
\usepackage{jheppub}
\usepackage{appendix}
\usepackage{mathtools}
\usepackage{orcidlink}
\usepackage{siunitx}
\usepackage{physics}
\usepackage{todonotes}
\usepackage{comment}
\usepackage{perpage}
\MakePerPage{footnote}
\usepackage{fontawesome5}

\newcommand{\ogw}{\Omega_\text{GW}}

\newcommand{\gs}{g_\star}

\newcommand{\Trh}{T_\text{rh}}

\newcommand{\mpl}{m_{\text{Pl}}}

\newcommand{\lrfrac}[2]{\left( \frac{#1}{#2} \right)}

\title{
Graviton Production from Inflaton Condensate: Boltzmann vs Bogoliubov
}
\author[a]{Chenhuan Wang\,\orcidlink{0009-0005-0858-050X}}
\author[b]{,Yong Xu\,\orcidlink{0000-0002-4582-8747}}
\author[c,d]{,Wenbin Zhao\,\orcidlink{0000-0002-1143-9989}}
\affiliation[a]{Bethe Center for Theoretical Physics and Physikalisches Institut\\ Universität Bonn,
Nussallee 12, 53115 Bonn, Germany}
\affiliation[b]{McGill University Department of Physics \& Trottier Space Institute\\
3600 Rue University, Montréal, QC, H3A 2T8, Canada}
\affiliation[c]{International Centre for Theoretical Physics Asia-Pacific, University of Chinese Academy of Sciences,
100190 Beĳing, China}
\affiliation[d]{Taiji Laboratory for Gravitational Wave Universe (Beijing/Hangzhou), University of Chinese Academy of Sciences (UCAS), Beijing, China.}
\emailAdd{cwang1@uni-bonn.de}
\emailAdd{yong.xu6@mcgill.ca}
\emailAdd{zhaowenbin@ucas.ac.cn}

\abstract{We study graviton production from an oscillating inflaton condensate during reheating by systematically comparing Boltzmann and Bogoliubov descriptions for inflaton potentials of the form $V(\phi)\propto\phi^n$ around the minimum. The Bogoliubov framework provides a unified description of graviton production, capturing both perturbative and non-perturbative effects across short and long wavelengths, whereas the Boltzmann approach is restricted to perturbative production at short wavelengths. For the quadratic case ($n=2$), we find that the two approaches yield identical graviton spectra at short wavelengths, indicating that the Boltzmann treatments fully captures perturbative gravitational production in this regime. For steeper potentials ($n>2$), however, we identify a sizable contribution arising from the non-adiabatic transition between inflation and reheating. This component is naturally incorporated in the Bogoliubov formalism but absent in the Boltzmann description, and we show that it is important over a broad range of momenta. We derive analytic approximations within both frameworks that clarify the physical origin and scaling behavior of the spectrum. Our results delineate the regime of validity of Boltzmann approaches and show that, for steeper inflaton potentials, graviton production is governed by non-adiabatic transition dynamics for which the Bogoliubov formalism provides the most appropriate description.}
\begin{document}
\begin{flushright}
April 2026
\end{flushright}
\maketitle

%%%%%%%%%%%%%%%%%%%%%%%%%%%%%%%%%%%%%
\section{Introduction}
%%%%%%%%%%%%%%%%%%%%%%%%%%%%%%%%%%%%%

Cosmic inflation offers a compelling framework for addressing key challenges in the standard cosmology \cite{Starobinsky:1980te, Guth:1980zm, Linde:1981mu, Albrecht:1982wi}.  In the simplest scenario, a single scalar field, the inflaton field, undergoes slow-roll evolution along a nearly flat potential, driving an era of exponential expansion. During this phase, primordial tensor perturbations are generated and subsequently stretched to superhorizon scales by the accelerated expansion. Once the inflation ends and the tensor modes reenter the horizon, they become dynamical, giving rise to a stochastic background of primordial gravitational waves (GWs) \cite{Starobinsky:1979ty, Allen:1987bk, Sahni:1990tx, Turner:1993vb}. The rapid expansion during inflation stretches the wavelengths of these modes, shifting them to very low frequencies. The amplitude of this GW background is directly tied to the energy scale of inflation, making it a powerful probe of the very early Universe \cite{Caprini:2018mtu, Giovannini:2019oii}.

Beyond this well-studied inflationary GW background, additional sources of gravitational radiation can arise during the post-inflationary evolution. In particular, during reheating the inflaton typically undergoes coherent oscillations around the minimum of its potential, inducing rapid time variations of the spacetime background. This non-adiabatic dynamics can excite gravitons whose physical wavelengths remain well inside the horizon throughout reheating. As a result, reheating generically produces a high-frequency GW component, distinct from the inflationary tensor spectrum.

The production of such high-frequency gravitons may be viewed as a manifestation of particle production in a time-dependent background. Two main approaches have been developed in the literature to compute this effect. A natural approach is the Bogoliubov formalism\footnote{We note that the Bogoliubov formalism has been widely used in the literature on non-perturbative particle production during (p)reheating \cite{Traschen:1990sw,Shtanov:1994ce,Kofman:1997yn,Brandenberger:2026trv}.}, which directly tracks the evolution of graviton mode functions and their associated Bogoliubov coefficients; see, for example, Refs.~\cite{Ema:2015dka,Ema:2016hlw,Ema:2020ggo, Alsarraj:2021yve, Pi:2024kpw, Giovannini:2025obx,Wang:2026pff} in  the context of graviton production. This mechanism is also known as gravitational particle production (GPP);  see Ref.~\cite{Kolb:2023ydq} for a comprehensive review. An alternative framework is based on the Boltzmann description, in which the graviton production is encoded in effective, local-in-time production rates sourced by the oscillating background. This production is governed by the phase-space Boltzmann equation, which has been developed and applied to graviton production in Refs.~\cite{Choi:2024ilx, Xu:2024fjl, Bernal:2025lxp,Xu:2025wjq,Datta:2025wfh}.

A systematic comparison between these two approaches is necessary for a reliable understanding of particle production during the inflaton oscillation phase. Similar comparative studies have recently been carried out for spin-$0$ scalar particles \cite{Kaneta:2022gug, Basso:2022tpd, Chakraborty:2025zgx}, as well as for fermions of spin-$1/2$ \cite{Bhusal:2025oqg} and spin-$3/2$ \cite{Kolb:2025wyj}. In this work, we extend this program to massless spin-$2$ gravitons. We go beyond the commonly studied quadratic case in the literature and consider a general class of reheating scenarios in which the inflaton oscillates in a power-law potential, $V(\phi) \propto \phi^n$ with $n \geq 2$ around the minimum. Within this setup, we analyze the graviton production using both the Bogoliubov\footnote{Related study, Ref.~\cite{Mudrunka:2026kgm}, using the Bogoliubov approach for general inflaton potentials has recently appeared, based on numerical simulations. In this work, we focus primarily on a systematic comparison between the Bogoliubov and Boltzmann approaches. In addition, we provide analytical approximations wherever possible to complement and elucidate the numerical results.} and Boltzmann frameworks. 

The primary goal of this paper is to provide a systematic and quantitative comparison of the Bogoliubov and Boltzmann approaches to graviton production during reheating, identifying the regimes where they agree and clarifying the origin of any discrepancies. In addition, we derive analytical approximations within both formalisms, allowing for a transparent physical interpretation of the underlying production mechanisms.

This paper is organized as follows. In Section~\ref{sec:setup}, we introduce the background dynamics and notation. The Boltzmann approach to graviton production is presented in Section~\ref{sec:boltz}, while Section~\ref{sec:bogo} is devoted to the Bogoliubov formalism. A detailed comparison of the two methods is carried out in Section~\ref{sec:comparison}. In Section~\ref{sec:gw_sp}, we compute the  gravitational wave spectrum. Our results are summarized in Section~\ref{sec:conclusion}.

In this work, $p$ is always used for physical momentum and the comoving momentum is given as $k = p(t) a(t)$. Reduced Planck mass $\mpl \equiv 1/\sqrt{8 \pi\, G} \simeq \SI{2.4e18}{\giga\eV}$ is used as the unit of energy. We adopt the metric signature $\eta_{\mu\nu} = \text{diag}(+, -, -, -)$. We acknowledge the use of the following programs: \cite{harris2020array, Hunter:2007, rackauckas2017differentialequations, Ellis_2017, Shtabovenko_2025}. 
Our numerical code, together with the Mathematica notebook for the matrix element calculations, is publicly available at \href{https://github.com/not-physicist/GPP_Bogo_v_Boltz}{\faGithubSquare}.

\section{The Setup}
\label{sec:setup}
We consider a minimal setup with the action
\begin{align}
S \supset \int d^4 x\sqrt{-g} \left[ \frac{\mpl^2}{2} R + \frac{1}{2} g_{\mu \nu}\partial^\mu \phi \partial^\nu \phi -V(\phi)\right]\,,
\label{eq:R-action}
\end{align}
where  $g$ denotes the determinant of the metric $g_{\mu\nu}$ and $R$ the Ricci scalar. The term $V(\phi)$ is the inflaton potential; during the inflaton oscillation it can be approximated as 
\begin{align}\label{eq:vphi}
V(\phi) \simeq \lambda \frac{\phi^n}{\mpl^{n-4}}\,,
\end{align}
where $\lambda$ is a dimensionless parameter; it can be fixed by the scalar power spectrum once a specific inflation model is considered. For $n = 2$, Eq.~\eqref{eq:vphi} corresponds to a quadratic potential with $m_\phi = \sqrt{2 \lambda}\, \mpl $, which arises from various inflationary models, including Starobinsky inflation~\cite{Starobinsky:1980te}, certain classes of $\alpha$-attractor models~\cite{Kallosh:2013hoa, Kallosh:2013maa}, and both small- and large-field polynomial inflation scenarios~\cite{Drees:2021wgd, Drees:2022aea}. Moreover, $\alpha$-attractor models~\cite{Kallosh:2013hoa, Kallosh:2013maa} can also lead to scenarios with $n > 2$. Throughout this work we will assume $\alpha$ attractor T model, see Appendix~\ref{app:T_model} for details on the model.

The usual Friedmann–Lemaître–Robertson–Walker (FLRW) metric suffices for solving the background
\begin{equation}
    \dd{s}^2 = \dd{t}^2 - a(t)^2 \delta_{ij}\dd{x}^i \dd{x}^j ,
\end{equation}
where the scale factor $a(t)$ characterizes the cosmic expansion. 
The classical homogeneous inflaton field $\phi \equiv \phi(t)$ follows its equation of motion and we have also the Friedmann equation and Boltzmann equation for the radiation energy density \cite{Turner:1983he, Garcia:2020wiy}
\begin{align}\label{eq:background}
\begin{split}
    \ddot{\phi} + (3 H + \Gamma_\phi) \dot{\phi} + V'(\phi) = 0, \, \\
    3 \mpl^2 H^2 =  \rho_\phi + \rho_r, \,
    \\
    \dot{\rho}_r + 4 H \rho_r = (1+\omega)\Gamma_\phi \rho_\phi,
    \end{split}
\end{align}
where $H \equiv \dot{a}/a$ is the Hubble parameter, $\rho_\phi \equiv \dot{\phi}^2/2 + V(\phi)$ denotes the inflaton energy density, and $\rho_r \equiv \gs\,\pi^2\, T^4/30$ represents the radiation energy density. Moreover, $\Gamma_\phi$ corresponds to the decay rate of inflaton, and the specific form will be discussed in more detail in Section \ref{sec:comparison}; $\omega \equiv p_\phi/\rho_\phi$ corresponds to the equation of state parameter with $p_\phi \equiv \dot{\phi}^2/2 - V(\phi)$ being the pressure density. We use a dot to denote derivative with respect to the cosmic time $t$, and for the inflation potential $V'(\phi) \equiv dV(\phi)/d\phi$. Here, $\gs$ accounts for the total number of relativistic degrees of freedom contributing to the radiation energy density. In Eq.~\eqref{eq:background}, the maximal value in the Standard Model $g_* = 106.75$ is used, as we assume a scale well above the electroweak phase transition during reheating. These background equations can only be solved numerically in general and some numerical details can be found in Appendix~\ref{app:num}.

The choice of the power $n$ in Eq.~\eqref{eq:vphi} determines the shape of the inflaton potential near its minimum and therefore controls the properties of the inflaton oscillations during reheating.  For a quadratic potential with $n=2$, the inflaton behaves as a damped harmonic oscillator with a constant mass $m_\phi$. In contrast, for $n\ge 4$ the inflaton oscillations are intrinsically anharmonic: the oscillation frequency depends on the oscillation amplitude and therefore evolves in time as the Universe expands. This qualitative difference plays an important role in determining the dynamics of reheating and the associated production of gravitational waves.

For the class of potentials in Eq.~\eqref{eq:vphi}, the time-averaged equation-of-state parameter of the oscillating inflaton field is given by $\omega =(n-2)/(n+2)$~\cite{Turner:1983he,Shtanov:1994ce}. When the energy density is dominated by the coherently oscillating inflaton, the inflaton
energy density redshifts as $\rho_\phi \propto a^{-3(1+\omega)}$. Since
$\rho_\phi \propto \Phi (t) ^n $, where $\Phi(t)$ denotes the oscillation amplitude, this implies $\Phi(t) \propto a^{-6/(n+2)}$. The time dependence of the oscillation amplitude directly translates into a time-dependent
oscillation frequency. Defining $\tilde m(t)$ as the characteristic oscillation frequency
of the inflaton condensate, which is set by the curvature of the potential at the
oscillation amplitude, one finds~\cite{Figueroa:2020rrl}
\begin{equation}
    \tilde m^2(t)
    \simeq
    \left.\frac{d^2V}{d\phi^2}\right|_{\phi=\Phi}
    \propto
    \Phi(t)^{\,n-2}
    \propto
    a(t)^{-6\omega}.
\label{eq:m_tilde_a}
\end{equation}

For $n=2$, this reduces to a constant oscillation frequency, while for $n>2$ the frequency redshifts with the expansion. This time dependence of $\tilde m$ is a key ingredient in the subsequent analysis of graviton production during reheating, as it governs the structure of resonant particle production for steeper inflaton potentials.

%%%%%%%%%%%%%%%%%%%%%%%%%%%
\section{Graviton Production from Inflaton Condensate - Boltzmann Method}
%%%%%%%%%%%%%%%%%%%%%%%%%%%
\label{sec:boltz}

In this section, we investigate the Boltzmann approach to computing the graviton
spectrum during reheating. The  spectrum has previously been
analyzed at the level of the integrated graviton energy density for $n\ge 2$ in
Ref.~\cite{Choi:2024ilx}. More recently, Refs.~\cite{Xu:2024fjl,Bernal:2025lxp,Xu:2025wjq}
studied the graviton spectrum at the level of the graviton number density and the
phase-space distribution function for the quadratic case $n=2$, modeling graviton
production as pair annihilation of inflaton, $\phi\phi\to hh$.

The main goal in this section is to extend these analyzes to general inflaton potentials
with $n>2$ and to derive the corresponding graviton phase space distribution
function. Different from previous treatments based on inflaton quanta, we model
the inflaton as a coherently oscillating classical background that acts as a source
for graviton production. This formulation allows us to capture the effects of
anharmonic inflaton oscillations and provides a unified framework applicable to
arbitrary $n$. These constitute the main new results presented in this section.

\subsection{Formalism}
We begin by defining the graviton field through a perturbative expansion of the
metric around flat spacetime
\begin{equation}
    g_{\mu\nu} = \eta_{\mu\nu} + \kappa\, h_{\mu\nu},
\label{eq:g-h-decomp}
\end{equation}
with $\kappa \equiv 2/\mpl$.    The field $h_{\mu\nu}$ represents the dynamical graviton degrees of freedom propagating on the Minkowski background. Here the metric is expanded around Minkowski spacetime because the collision term describes graviton emission as a local process, for which the microscopic interaction timescale is much shorter than the Hubble timescale. The cosmological expansion is then incorporated separately through the Boltzmann evolution, which will be discussed below. In this section, we refer to $h_{\mu\nu}$ as the graviton field. As for any other particle species, in an expanding Universe the phase-space distribution function of gravitons, $f_h$, is governed by the phase space  Boltzmann equation
\cite{Kolb:1990vq}:

\begin{equation} \label{eq:fh-boltz}
    \frac{\partial f_h}{\partial t} - H\, p\, \frac{\partial f_h}{\partial p} =  \mathcal{C}_h\,,
\end{equation}
where $\mathcal{C}_h$ denotes the collision term to the graviton production and $p$ its physical momentum. When the inflaton is treated as a classical source, the collision term describing the
production of a graviton pair with physical four-momenta $p$ and $p'$
can be written as
\begin{align}\label{eq:Ch}
    \mathcal{C}_h (a, p) &= \frac{1}{2E}\sum^{+\infty}_{j=-\infty} \int \frac{g_h}{(2\pi)^3 }\frac{ \dd[3]{p'}}{2 E'} (2\pi)^4 |\mathcal{M}_j|^2 \delta^{(4)}(p_j - p - p')\,.
\end{align}
Here we introduce $p_j \equiv (2\tilde mj, 0)$ to denote the effective
four-momentum associated with the inflaton condensate, where the integer $j$
labels the Fourier modes of the oscillating background, to which we return
below. $\mathcal{M}_j$ denotes the matrix element for the corresponding Fourier mode, which we shall specify in our case later. 

Eq.~\eqref{eq:fh-boltz} can be conveniently solved by writing into a total derivative and working in terms of the
comoving momentum $k$, in which case the Hubble expansion term can be absorbed,
reducing the Boltzmann equation to an ordinary differential equation:
\begin{equation} \label{eq:fh-boltz2}
    \frac{d f_h (a, k)}{d a} =  \frac{1}{a\, H}\mathcal{C}_h\,.
\end{equation}
The solution of Eq.~\eqref{eq:fh-boltz2} is written as an integral over the scale factor
\begin{equation} \label{eq:fh-from-int}
    f_h(a, k) = \int_{a_e}^a \frac{da'}{a'\, H(a')}\, \ \mathcal{C}_h\left(a', \frac{k}{a'}\right)\,.
\end{equation}
We assume that at the end of inflation ($a=a_e$), i.e. at the onset of reheating, no gravitons are present in the momentum range of interest. This assumption will be further examined in Section~\ref{sec:comparison} and Appendix~\ref{app:dominance}.

Our next goal is to compute the collision term $\mathcal{C}_h$ in Eq.~\eqref{eq:Ch}. To this end, we first need the matrix element $\mathcal{M}_j$. In the following, we briefly revisit the relevant steps of the calculation. We begin by deriving the interaction terms between gravitons and the background inflaton field.  
By expanding the Einstein-Hilbert action (the first term in Eq.~\eqref{eq:R-action}) with Eq.~\eqref{eq:g-h-decomp}, we can get the graviton propagator and triple graviton vertex. The scalar-graviton interaction can be obtained by expanding the Lagrangian for inflaton field $\sqrt{-g} \mathcal{L}_\phi$, which yields the following interaction terms up to second order in $\kappa$~\cite{Holstein:2006bh} 
\begin{subequations}
\begin{align}
\sqrt{-g} \mathcal{L}_\phi^{(0)} & =\frac{1}{2}\partial_\mu \phi \partial^\mu \phi-V(\phi)\,, \label{eq:Lphi0} \\
\sqrt{-g} \mathcal{L}_\phi^{(1)} & =-\frac{\kappa}{2} h^{ \mu \nu}\left[\partial_\mu \phi \partial_\nu \phi- \eta_{\mu \nu}\left(\frac{1}{2}\partial_\alpha \phi \partial^\alpha \phi-V(\phi)\right)\right]\,,\label{eq:Lphi1}\\
\sqrt{-g} \mathcal{L}_\phi^{(2)} & =\frac{\kappa^2}{2}\left(h^{\mu \lambda} h^{\nu}{ }_\lambda-\frac{1}{2} h h^{ \mu \nu}\right) \partial_\mu \phi \partial_\nu \phi \nonumber \\
& \quad -\frac{\kappa^2}{4}\left(h^{\alpha \beta} h_{\alpha \beta}-\frac{1}{2} h^{2}\right)\left(\frac{1}{2} \partial^\alpha \phi \partial_\alpha \phi-V(\phi)\right) \label{eq:Lphi2}\,,
\end{align}
\end{subequations}
where $h$ denotes the trace of graviton field. 

When dealing with on-shell gravitons, the above interaction terms can be simplified using the transverse and traceless (TT) gauge of graviton with the following constraints
\begin{equation}\label{eq:TT}
    \partial_\mu h^{\mu \nu} = 0\,, \quad h=h^{\mu}_{\mu}=h^{\mu\nu} \eta_{\mu\nu}=0\,,\quad h^{0\mu}=0\,.
\end{equation}

The inflaton field here corresponds to a homogeneous condensate and is therefore treated as a time-dependent background for the graviton production; consequently, all spatial derivatives vanish, $\partial_i \phi = 0$. Thus, the first term within the bracket of Eq.~\eqref{eq:Lphi1} does not contribute if the graviton is on-shell; the remaining non-vanishing terms within the bracket reduce to $\dot{\phi}^2/2 - V(\phi)$. These two quantities are time-dependent and this dependence enters the time integral for the transition amplitude. In what follows, we assume that the inflaton oscillation frequency is much larger than both the Hubble expansion rate and the inflaton decay rate, $\tilde m \gg H, \Gamma_\phi$, so that the inflaton undergoes many coherent oscillations within a Hubble time. We decompose the kinetic term $K$ and potential term $V$ into ``Fourier series'' as follows \footnote{Note that cosmic time $t$ in this context is counted from the end of inflation $t_e$, i.e.~$t \to t-t_e$, which is implied onward.}
\begin{subequations}\label{eq:K-V-four-decomp}
\begin{align}
    K_\phi(t) &= \frac{1}{2}\dot\phi^2(t) = \rho_\phi (t) \sum_{j=-\infty}^\infty K_j e^{-2i\tilde{m}jt},\label{eq:K-four-decomp} \\
    V_\phi(t) &= V(\phi(t)) = \rho_\phi(t) \sum_{j=-\infty}^\infty V_j e^{-2i\tilde{m}jt},\label{eq:V-four-decomp}
\end{align}
\end{subequations}
where the factor $2$ in the exponential is introduced, so that $\tilde{m} = m_\phi$ for quadratic potential. $K_j$ and $V_j$ are both dimensionless Fourier coefficients. The inflaton field itself oscillates with the period $ T = 2\pi/\tilde m$. Kinetic and potential energies oscillate with half of the period $T/2$ (independent of the power of potential, as long as it is even, positive power).  The zeroth mode corresponds to the constant (over one oscillation) background and does not excite graviton here. Since $\tilde{m}=\tilde{m}(t)$ with $n \geq 4$ potential, such decomposition must be carried out for \textit{each} oscillation. Then the Fourier coefficients $K_j, V_j$ can also change from oscillation to oscillation in general. 

\begin{figure}[ht!]
    \centering
    \includegraphics[width=0.18\linewidth]{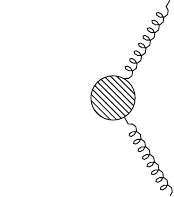}
    \qquad \qquad
    \includegraphics[width=0.2\linewidth]{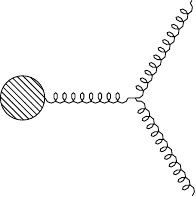}
    \caption{Two processes $\mathcal{M}_1$ and $\mathcal{M}_2$ contributing to the graviton production. The shaded blob represents the inflaton field $\phi(t)$ as a classical source.}
    \label{fig:graviton-proc}
\end{figure}

With the Lagrangian, we can compute the matrix elements of processes shown in Fig.~\ref{fig:graviton-proc} by writing out the Feynman rules. They are listed in Appendix \ref{sec:feynman-rules}. After combining both diagrams and their interference (shown in Eq.~\eqref{eq:mat-elements}), the kinetic energy contributions cancel out and we end up with:
\begin{equation}
    \frac{1}{4}\sum_{\sigma, \sigma'} |\mathcal{M}_{\text{tot}, j}|^2 = \frac{\rho_\phi^2(t)|V_j|^2}{2\mpl^4}\,, 
    \label{eq:tot-matrix-elem}
\end{equation}
where the factor $1/4$ arises from averaging over final state (two gravitons) polarizations.

The collision term defined in Eq.~\eqref{eq:Ch} now can be expressed as
\begin{align}\label{eq:Ch_res}
    \mathcal{C}_h (a, k/a) &= 
    \frac{2\pi g_h}{2E} \sum_{j=-\infty}^{+\infty} \int \frac{\dd[3]{p'}}{2E'} \frac{1}{4} \sum_{\sigma, \sigma'}|\mathcal{M}_{\text{tot}, j}|^2 \delta(2\tilde{m}j - E -E') \delta^{(3)} (-\vec{p} - \vec{p}'), \notag \\
     &= \sum_{j=-\infty}^{+\infty} \frac{\pi}{4p^2} \frac{ \rho_\phi^2 (t) |V_j|^2}{\mpl^4} \delta(\tilde{m}j  - p)\,,
\end{align}
where we have used $g_h=2$ for on-shell graviton. The Dirac delta enforces the value of scale factor in the final spectrum with $a=a_k$, where $a_k$ is the solution to the implicit equation
\begin{equation}\label{eq:position_Bol}
    a_k \tilde{m}(a_k) j = k.
\end{equation}
It leads to lower/upper bound on the comoving momentum $k$ depending on $\tilde{m}(a)$, since we only consider such graviton production after the end of inflation in Eq.~\eqref{eq:fh-from-int}. In potentials with $n > 4 $ considered here, the oscillation frequency generally decreases, leading to an upper bound on $k$. It is also clear from Eq.~\eqref{eq:position_Bol} that only $j>0$ modes contribute to the graviton production.

\subsection{Solution for Graviton Phase Space Distribution Function} 
With the collision term given in Eq.~\eqref{eq:Ch_res}, we are now ready to solve the Boltzmann Eq.~\eqref{eq:fh-boltz2} and obtain the graviton distribution function $f_h(k)$. We first consider the simplest case of a quadratic inflaton potential, corresponding to $n=2$ in Eq.~\eqref{eq:vphi}, and subsequently generalize the discussion to $n>2$.

For the scenario where $\tilde{m} = \text{const.}$, Eq.~\eqref{eq:fh-from-int} gives
\begin{equation}
    f_h \left(a, \frac{k}{a} \right) = \frac{\pi}{4\mpl^4} \sum_{j>0} \frac{1}{\tilde{m}^3 j^3} \left(\frac{\rho_\phi^2 |V_j|^2}{H} \right)_{a=a_k} \Theta(k - a_e \tilde{m} j)\,.
    \label{eq:exact_boltzmann}
\end{equation}
The Heaviside function comes from the Dirac delta along with a lower boundary of the integral. For a quadratic inflaton potential, $V(\phi) = m_\phi^2 \phi^2/2$. We can write $\phi(t) \propto \sin(m_\phi t)$,
which gives $V_{\pm 1} = -1/4$, and $V_{|j|\ge 2} = 0$. Since only modes $j>0$ contribute, we have
\begin{equation}
    f_h(a, k) = \frac{\pi}{64} \frac{1}{m_\phi^3 \mpl^4} \left . \left( \frac{\rho_\phi^2}{H} \right) \right|_{a/a_e=k/a_e m_\phi} \Theta (k - a_e m_\phi)\,.
    \label{eq:boltzmann-ana}
\end{equation}
By writing $\rho_\phi \simeq 3\mpl^2 H^2 
\simeq 3\mpl^2 H_e^2 (a/a_e)^{-3} \exp(-2\Gamma_\phi/ 3 H) $, Eq.~\eqref{eq:boltzmann-ana} becomes
\footnote{
This result is consistent with the phase-space distribution obtained in the particle-scattering picture \cite{Xu:2025wjq} up to an overall factor of two.
}
\begin{equation}
    f_h(a, k) = \frac{9\pi}{64} \lrfrac{H_e}{m_\phi}^3 \lrfrac{k}{a_e m_\phi}^{-9/2} \exp \left(-\frac{2\Gamma_\phi}{H_e} \lrfrac{k}{a_em_\phi}^{3/2} \right) \Theta (k - a_e m_\phi)\,.
    \label{eq:boltzmann-ana-2}
\end{equation}
Eq.~\eqref{eq:position_Bol} tells us that graviton with large comoving momentum is produced late. The exponential appears because the inflaton field acts as the energy source for graviton production and the inflaton field experiences the exponential decay at the production time of graviton with large $k$.  

% % % % % % 
Now in general, $\tilde{m} = \tilde{m}(a)$ can be time-dependent, especially in the case of $n>2$ monomial potential. In this case, we obtain the spectrum
\begin{equation}
    f_h (k) = \frac{\pi}{4\mpl^4} \sum_{j>0} \left[ \frac{ C_j(a)}{\tilde{m}^3 j^3} \lrfrac{\rho_\phi^2 |V_j|^2}{H} \right]_{a=a_k} \Theta(a_k-a_e).
    \label{eq:boltzmann-exact}
\end{equation}
With $n=4$, the spectrum diverges with a series of Dirac delta's. Using Eq.~\eqref{eq:boltzmann-exact} to compute the phase space distribution numerically automatically incorporates the Heaviside function or Dirac deltas, as $\tilde{m}$ is only meaningful during reheating. The correction factor $C_j(a)$ coming from the time-dependence of the oscillation frequency, defined as such
\begin{equation}
    C_j(a)^{-1} = \left|1 + \frac{k}{j} \frac{d\tilde{m}(a)/da}{\tilde{m}(a)^2} \right| = \left|1 + \frac{1}{j}\frac{k}{a_e H_e} \frac{a_e H_e}{aH} \frac{d\tilde{m}/dt}{\tilde{m}^2} \right|.
    \label{eq:boltzmann-correction}
\end{equation}
If one follows the scaling in Eq.~\eqref{eq:m_tilde_a}, it becomes universal for all $j$
\begin{equation}
    C^{-1} (a_k) 
    % = \left| 1 - \frac{3(n-2)}{n+2} \right|
    =2 \left|\frac{n-4}{n+2} \right|.
    \label{eq:boltzmann-correction2}
\end{equation}

Furthermore, the background can be approximated as $\rho_\phi \propto H^2 \propto a^{-3(1+\omega)}$. Assuming the Fourier coefficients can be taken to be constant, the phase space distribution for $n>4$ reads:
\begin{equation}
    f_h = \frac{9\pi}{8} \frac{n+2}{|n-4|} \sum_{j>0} \lrfrac{H_e}{\tilde{m}_e j}^3 \lrfrac{k}{a_e \tilde{m}_e j}^{\frac{9}{4}\frac{4}{n-4}} |V_j|^2 \Theta(a_e \tilde{m}_e j - k)\,,
    \label{eq:boltzmann-power}
\end{equation}
with $\tilde{m}_e = \tilde{m}(a_e)$. In this simple analytical expression, we omit the factor associated with the inflaton decay into daughter particles: such decay is complicated with $n\geq4$ and is not a simple exponential decay~\cite{Garcia_2021}. We note that Eq.~\eqref{eq:boltzmann-power} leads to a GW spectrum  
% $\propto k^{(4n-7)/(n-4)}$ for each mode just as shown as 
consistent with Eq.~(45) in Ref.~\cite{Choi:2024ilx} and Eq.~(35) in Ref.~\cite{Mudrunka:2026kgm}. As will be shown in Section \ref{sec:comparison}, this scaling \textit{does not} determine the envelope of the whole (Boltzmann) spectrum, which is instead determined primarily by the Fourier coefficients. The correction factor diverges for quartic potential and it leads to the $1/|n-4|$ factor. This can be traced back to the Dirac delta in Eq.~\eqref{eq:Ch_res} with $\tilde{m}(a) \propto a^{-1}$. This perfect scaling is however not observed in numerical solutions, so the general formula in Eq.~\eqref{eq:boltzmann-exact} still delivers sensible results.

%%%%%%%%%%%%%%%%%%%%%%%%%%%%%%%%%%%%%%%%%%%%%%%%%%%%%%%%%%%%%%%%%%
\section{Gravitational Particle Production - Bogoliubov Method} \label{sec:bogo}
%%%%%%%%%%%%%%%%%%%%%%%%%%%%%%%%%%%%%%%%%%%%%%%%%%%%%%%%%%%%%%%%%%
In this section, we take a different approach on post-inflationary graviton production, namely through the gravitational particle production (GPP) \cite{Kolb:2023ydq}, which is also referred as the Bogoliubov method. As in the previous section, we first revisit the basic formalism and then present two general analytical approximations for the Bogoliubov coefficients before and during inflaton oscillations. Wherever possible, we compare our results with those in the existing literature.

\subsection{Formalism}
Here, we consider a slightly different perturbative expansion of the metric \footnote{Note that the tensor perturbation here is defined differently compared to in Eq.~\eqref{eq:g-h-decomp}. There the decomposition takes place in Minkowski background and the the tensor field has non-zero mass dimension.} 
\begin{align}
ds^2  & = dt^2 - a^2(t) \left(\delta_{ij} +  \tilde{h}_{ij}\right) dx^i dx^j \nonumber \\
& = a^2(\tau) \left[d\tau^2 - \left(\delta_{ij} + \tilde{h}_{ij}\right) dx^i dx^j\right]\,,
\label{eq:metric}
\end{align}
where we have used the conformal time $\tau  = \int dt/a(t)$. Then, one can obtain the action for the graviton field from the linearized Einstein-Hilbert action \cite{Ema:2020ggo, gorbunovIntroductionTheoryEarly2011}:
\begin{align}
S &= \frac{\mpl^2}{8} \int \dd{t} \dd[3]{x} a^3  \left[ \left(\dot{\tilde{h}}_{ij}\right)^2 - \frac{1}{a^2} \left(\partial_k \tilde{h}_{ij}\right)^2\right], \nonumber \\
% & = \sum_{\lambda = +\,, \times } \int \frac{\dd{\tau} \dd[3]{k}}{(2\pi)^3} \frac{1}{2} \left[ |\tilde{h}_{\lambda}^{\prime}(k)|^2 -\omega^2_k |\tilde{h}_{\lambda}(k)|^2\right]\,,
& = \sum_{\sigma = +\,, \times } \int \frac{\dd{\tau} \dd[3]{k}}{(2\pi)^3} \frac{1}{2} \left[ |{h}_{k}^{\prime}(\tau, \sigma)|^2 -\omega^2_k |{h}_{k}(\tau, \sigma)|^2\right]\,,
\label{eq:bogo_action}
\end{align}
where $\sigma$ denotes the polarization of the graviton. Direct coupling of graviton and inflaton field is not considered here. We have written the tensor perturbation $\tilde{h}_{ij}$ into such decomposition:
% We write the tensor perturbation as a quantum field with the following decomposition
\cite{Kolb:2023ydq, gorbunovIntroductionTheoryEarly2011}
\begin{equation}
    \hat{\tilde{h}}_{ij}(\tau, \vec{x}) = \frac{1}{a} \frac{2}{\mpl} \sum_{\sigma=+, \times} \int \frac{\dd[3]{k}}{(2\pi)^3} \left( e^{i\vec{k}\cdot \vec{x}} h^*_k(\tau, \sigma) \hat{a}_{k, \sigma} + e^{-i\vec{k}\cdot\vec{x}} h_k(\tau, \sigma) \hat{a}^\dagger_{k, \sigma} \right) \epsilon_{ij}^\sigma\,.
    \label{eq:h-decomp}
\end{equation}
Here, $\hat{a}$ and $\hat{a}^\dagger$ are the usual ladder operators
% with the usual algebra
, and $\epsilon_{ij}^\lambda$ represents the graviton polarization tensor. Moreover, we have defined the effective frequency as
\begin{equation}
    \omega_k^2(\tau) \equiv k^2 -\frac{a^2(\tau) R(\tau)}{6} = k^2 - \frac{a''(\tau)}{a(\tau)}\,.
    \label{eq:effective_frequency}
\end{equation}

It turns out that although the above decomposition can be done at all times, it is not unique. This comes from the variability of the vacuum state in curved spacetime. The labels ``IN'' and ``OUT'' denote the quantities taken at infinite past/future. They are related by the Bogoliubov coefficients $\alpha_k, \beta_k$ \cite{Kolb:2023ydq}
\begin{align}\label{eq:Bogo_coefficients}
\begin{split}
    h^{\text{IN}}_k(\tau) &= \alpha_k h_k^\text{OUT}( \tau) + \beta_k h_k^{\text{OUT}, *}(\tau), \\
    \hat{a}^\text{IN}_k &= \alpha^*_k \hat{a}^{\text{OUT}}_k -  \beta^*_k \hat{a}^{\text{OUT}, \dagger}_{-k}\,,
\end{split}
\end{align}
with the constraint $|\alpha_k|^2 -|\beta_k|^2 =1$ for bosons.

% With the help of the Wronskian condition $h h'^{*} - h^* h' = i$, we can calculate the Bogoliubov coefficients up to time dependent phase factors through the ``IN'' and ``OUT'' mode functions~\cite{Kolb:2023ydq}:  \cw{We don't need this equation anymore, right?}
% \begin{equation}
% \begin{aligned}\label{eq:Bogo_from_modef}
% & \alpha_k=i\left[h^{\text{OUT}, *}_k(\tau) h^{\text{IN}\prime}_k(\tau)-h^{\text{OUT}\prime, *}_k(\tau) h^{\text{IN}}_k(\tau)\right]\,, \\
% & 
% \beta_k=i \left[h^{\text{OUT}\prime}_k(\tau) h^{\text{IN}}_k(\tau)-h^{\text{OUT}}_k(\tau) h^{\text{IN}\prime}_k(\tau)\right]\,.
% \end{aligned}
% \end{equation}

The total particle number, as viewed by an observer in the infinite future, can be calculated by taking the vacuum expectation value of the number operator $\hat{N} = (2\pi)^{-3} \int \hat{a}_k^{\mathrm{OUT}, \dagger} \hat{a}_k^{\mathrm{OUT}} \, \mathrm{d}^3 k.$
From this, the comoving number density 
per logarithmic interval (per degree of freedom) 
can be expressed as \cite{Kolb:2023ydq}
\begin{equation}
    a^3 n_k \equiv a^3\dv{ n}{\ln k} = \frac{k^3}{2\pi^2} |\beta_k|^2.
\end{equation}

Using the equation of motion for field operator $\tilde{h}_{ij}$ from Eq.~\eqref{eq:bogo_action}, we obtain a differential equation for the mode function
\begin{equation}
    h''_k(\tau) + \omega_k^2(\tau) h_k(\tau) = 0.
    \label{eq:EOM-h}
\end{equation}
This equation coincides with the equation governing GPP of scalar particles \cite{Kolb:2023ydq}. For the graviton mode functions in Eq.~\eqref{eq:EOM-h}, we impose the Bunch--Davies initial condition deep inside the horizon, which reads \cite{Kolb:2023ydq}
\begin{align}\label{eq:BD}
    \lim_{\tau \to -\infty }h_k(\tau) \simeq \frac{1}{\sqrt{2k}}\, e^{-i k \tau}, \qquad 
     \lim_{\tau \to -\infty } h_k'(\tau) \simeq -i k\, h_k(\tau),
\end{align}
where the prime represents a derivative with respect to $\tau$. These initial conditions are implemented at sufficiently early times such that $k \gg a H$, ensuring the validity of such initial state.

The solution to Eq.~\eqref{eq:EOM-h} can be parametrized with the Bogoliubov coefficients \cite{Kolb:2023ydq}
\begin{equation}\label{eq:hk}
h_k(\tau) = \frac{\alpha_k(\tau)}{\sqrt{2\omega_k(\tau)}} e^{-i \int^\tau \omega_k(\tau') \dd{\tau'}}
+ \frac{\beta_k(\tau)}{\sqrt{2\omega_k(\tau)}} e^{+i \int^\tau \omega_k(\tau') \dd{\tau'}}\,.
\end{equation}
We note that the initial condition Eq.~\eqref{eq:BD} corresponds to $\alpha_k \to 1$, and  $\beta_k  \to 0$. It follows from Eq.~\eqref{eq:hk} that
\begin{equation}
\omega_k h_k - i h_k' = \sqrt{\frac{\omega_k}{2}} \, 2 \beta_k e^{+i \int \omega_k d\tau} \,.
\end{equation}
which leads to \cite{Kolb:2023ydq, Kofman:1997yn}
\begin{equation}
|\beta_k|^2 = \frac{1}{2\omega_k} \left| \omega_k h_k - i h_k' \right|^2 \,.
\label{eq:bogo-exact}
\end{equation}
%
% \sout{To this end, we need to numerically solve Eq.~\eqref{eq:EOM-h} with an evolving background. Throughout this work, we utilize Eq.~\eqref{eq:bogo-exact} to obtain the exact numerical spectrum in the Bogoliubov approach.}

% Analytical approximation also exists. 
Using Eq.~\eqref{eq:hk} and  Eq.~\eqref{eq:EOM-h}, one obtains the differential equations of the Bogoliubov coefficients
\begin{equation}
\alpha_k' = \frac{\omega_k'}{2\omega_k} e^{2i \int^\tau \omega_k(\tau') \dd{\tau'}} \beta_k, \quad
\beta_k' = \frac{\omega_k'}{2\omega_k} e^{-2i \int^\tau \omega_k(\tau') \dd{\tau'}} \alpha_k.
\label{eq:EOM-alpha}
\end{equation}
After solving the background equations in Eq.~\eqref{eq:background}, Eqs.~\eqref{eq:EOM-h} or~\eqref{eq:EOM-alpha} are solved numerically to obtain the number density. Further details can be found in Appendix \ref{app:num}.

In the regime where $|\beta_k| \ll 1$,  one can take $\alpha_k \approx 1$; 
it then follows from Eq.~\eqref{eq:EOM-alpha} that  \cite{Kolb:2023ydq}
\begin{equation} 
    \beta_k \simeq \int_{\tau_i}^{\tau_f} \frac{\omega'_k}{2\omega_k} e^{-2i\int^\tau \omega_k(\tau') \dd{\tau'}} \dd{\tau}\,.
    \label{eq:bogo-approx}
\end{equation}
% % 
For short wavelength modes $k^2 \gg |a^{\prime \prime}/a|$, the mode frequency Eq.~\eqref{eq:effective_frequency} is well approximated by
$\omega_k\simeq k$. The integral over $\omega_k$ in the exponent for Eq.~\eqref{eq:bogo-approx} can be approximated by $k\tau$, i.e. $ \int^\tau \omega_k(\tau') \dd{\tau'} = \int^\tau \sqrt{k^2-a''/a}\dd{\tau'} \simeq k\tau\,$.  In this limit the Bogoliubov integral Eq.~\eqref{eq:bogo-approx} simplifies to be
\begin{align}
\beta_k & \simeq  \int_{\tau_i}^{\tau_f} \dd{\tau} \frac{\omega'_k}{2\omega_k} e^{-2i\int^\tau \omega_k(\tau') \dd{\tau'}}   \simeq \int_{\tau_i}^{\tau_f}   \dd{\tau} \frac{d(\omega^2_k)}{d\tau} \frac{1}{4\omega_k^2} e^{-2i k \tau},  \nonumber \\
&\simeq - \frac{1}{4k^2} \int^{\tau_f}_{\tau_i}  d\tau \, F'(\tau) e^{-2ik\tau},
\label{eq:beta_Fprime_app}
\end{align}
where $F(\tau) \equiv a^2 R/6$. The quantity $F(\tau)$ naturally can be split into a transition part and an oscillatory reheating part,  $F(\tau) = F_{\rm tr}(\tau) + F_{\rm rh}(\tau),$ where $F_{\rm tr}$ describes the background from inflation to the
oscillatory era around the end of inflation $\tau_e$, while $F_{\rm rh}$ contains the subsequent time-dependent reheating background during inflaton oscillation. Correspondingly, the Bogoliubov coefficients can be decomposed into a transition part and an oscillatory part, which we will investigate in the following two subsections. 
% The full Bogoliubov formalism automatically incorporates all these effects, and transition contribution. Therefore, when the oscillation contribution is dominant, it is expected that the Boltzmann and Bogoliubov approaches would yield identical results.
% \begin{equation}
% \beta_k = \beta_k^{\rm tr} + \beta_k^{\rm reh},
% \qquad
% \beta_k^{\rm tr} \equiv - \frac{1}{4k^2} \int d\tau \, F'_{\rm tr}(\tau)e^{-2ik\tau}.
% \label{eq:beta_transition_def_app}
% \end{equation}
% As mentioned previously, below Eq.~\eqref{eq:bogo-approx}, the reheating contribution can be decomposed further into a fast oscillatory
% piece and a slow envelope piece as discussed just below Eq.~\eqref{eq:a2R_der}. The former can be captured by the stationary-phase
% approximation as investigated in previous section. \cw{It might be more clear to split $F$ or $\beta$ in the beginning of 4.2, where we write out the slow and fast contribution.}

% While $\tau_f$ can be uncontroversially set as the conformal time at or after reheating, we find that using end of inflation as the lower limit $\tau_i = \tau_e$ in some cases gives incorrect/incomplete results. This is because the integral in Eq.~\eqref{eq:bogo-approx}, evaluated with $\tau_i=\tau_e$, does not account for graviton production associated with the non-adiabatic change of the background due to the transition from inflation to reheating at $\tau=\tau_e$. We will come back to this  in Section~\ref{sec:transition}.

Before closing this section, we note that for long-wavelength modes (small $k$), the mode function $h_k$ can exhibit non-perturbative behavior in regimes where $a^{\prime\prime}/a$ becomes negative in Eq.~\eqref{eq:effective_frequency}. Such non-perturbative graviton production has been studied in the literature; see, e.g., Ref.~\cite{Alsarraj:2021yve}. We will see this more explicitly using the numerical results presented in Section~\ref{sec:comparison}. Before that, we derive semi-analytical expressions based on Eq.~\eqref{eq:beta_Fprime_app} in the following subsections.

\subsection{Solution for Bogoliubov Coefficient: Oscillation Contribution}
In this subsection, we investigate the oscillation contribution, and present semi-analytical solutions to Eq.~\eqref{eq:beta_Fprime_app} during inflaton oscillation. 

To determine $F_\text{rh}$ in Eq.~\eqref{eq:beta_Fprime_app}, we require first $R$, which is given by 
\begin{align}
R& = \frac{\rho - 3 p}{\mpl^2} = \frac{4 V_\phi - 2 K_\phi}{\mpl^2} \,,
\end{align}
where the definition for the total energy  $\rho = \rho_\phi + \rho_r = V_\phi+ K_\phi + \rho_r$, and the total pressure density $p = p_\phi + p_r = K_\phi - V_\phi + p_r$  have been utilized. The radiation does not contribution due to cancellation. It follows that \textit{during reheating}, one can express $F(\tau)$ in terms of Fourier series with the help of Eq.~\eqref{eq:V-four-decomp}
\begin{align} \label{eq:a2R_new}
       F_\text{rh}(\tau) = \frac{a^2 R}{6} &=   
        \frac{a^2 \rho_\phi}{3\mpl^2}  \frac{2V_\phi-K_\phi}{\rho_\phi}  \nonumber \\
       & = - \frac{a^2 \rho_\phi}{3\mpl^2} \left( 1-3 \sum_{j=-\infty}^{+\infty} V_{j} e^{-2i\tilde{m} j t}\right)\,,
\end{align}
where we have used $K_\phi = \rho_\phi - V_\phi$. 

We now compute the conformal-time derivative of Eq.~\eqref{eq:a2R_new}. Using $\partial_\tau = d/d\tau = a\, d/dt$,  we  obtain 
\begin{align}\label{eq:a2R_der}
% \partial_\tau \lrfrac{a^2 R}{6} 
F'_\text{rh}(\tau)
&=
-\partial_\tau \lrfrac{a^2 \rho_\phi}{3\mpl^2} \cdot
\left(1-3\sum_{j=-\infty}^{+\infty} V_j e^{-2i\tilde m j t}\right) 
- \frac{a^2\rho_\phi}{3\mpl^2}
\partial_\tau \left(1-3\sum_{j=-\infty}^{+\infty} V_j e^{-2i\tilde m j t}\right) \nonumber \\
& =\frac{\partial_\tau (a^2\rho_\phi)}{a^2\rho_\phi}
\lrfrac{a^2 R}{6} + \frac{2i a^3\rho_\phi}{\mpl^2}(\tilde m+\dot{\tilde m}t)
\sum_{j=-\infty}^{+\infty} jV_j e^{-2i\tilde m j t}\,.
\end{align}
The rapid oscillations of the inflaton induce fast time variations in $\omega_k^2$, as seen from the second term of Eq.~\eqref{eq:a2R_der}, potentially leading to efficient graviton production. It arises from short-time-scale variation of the background and can be described in terms of instantaneous, local-in-time production rates similar to the Boltzmann formalism. Note that the first term also contains oscillatory factors, however this is negligible contribution compared to the other two contributions. As mentioned before $V_j$ can vary from oscillation to oscillation, but we find it is negligible $\dot{V}_j \ll \tilde{m}+\dot{\tilde{m}}t$.

For such oscillation contribution,  the Bogoliubov coefficient $\beta_k$ in Eq.~\eqref{eq:beta_Fprime_app}  can be expressed as 
\begin{align}\label{eq:beta_k_n}
     \beta_k^{\text{rh}}
     & = -\frac{1}{4 k^2} \int_{\tau_i}^{\tau_f}   \dd{\tau} F'_\text{rh}(\tau) e^{-2i k \tau},  \nonumber\\
      & \simeq \frac{i}{2k^2 \mpl^2} \sum_{j=-\infty}^{+\infty} \int_{\tau_i}^{\tau_f} \dd{\tau} a^3 \rho_\phi (\dot{\tilde{m}}t + \tilde{m}) j V_j^*  \exp \left[2i(j\tilde{m}t -k\tau)\right]\,,
\end{align}
where we have relabeled $-j \to j$ and the symmetry of Fourier coefficient in the last line. 

To evaluate Eq.~\eqref{eq:beta_k_n}, we apply the stationary phase approximation (SPA). To this end, we define the phase function in Eq.~\eqref{eq:beta_k_n} as 
\begin{equation}
    g(\tau) \equiv 2(j \tilde{m}t-k \tau),
\end{equation}
and the stationary phase approximation leads to
\begin{align}
    \beta_k^\text{rh} &
    \simeq \frac{i}{2k^2 \mpl^2} \sum_{j>0} \left( a^3 \rho_\phi (\dot{\tilde{m}}t + \tilde{m})jV^*_j e^{i\psi} \sqrt{\frac{2\pi}{ |g''|}} \right)_{a=a_s},   \notag \\
    &\simeq \frac{i}{2k \mpl^2} \sum_{j>0} a_s^2 \rho_\phi(a_s) \left( V^*_j e^{i\psi(\tau)} \sqrt{\frac{2\pi}{ |g''|}} \right)_{a=a_s} \,.
      \label{eq:bogo-ana}
\end{align}
The phase \textit{after} the application of the stationary point approximation is 
\begin{equation}
   \psi(\tau) \equiv 2(j\tilde{m}t-k\tau) + \text{sign}[g''(\tau)] \frac{\pi}{4} \, ,
\end{equation}
and the derivatives of the phase function $g(\tau)$ are
\begin{align}
\begin{split}
    g'(\tau) &= 2 j a \left(1+ t \partial_t \right)\tilde{m} -2k \simeq  \frac{4}{n}ja\tilde{m}-2k, \\
    g''(\tau) &= 2ja^2 \left(t \partial_t^2 + (2+Ht)\partial_t + H \right) \tilde{m}\simeq -8 \frac{n-4}{n(n+2)} j a^2 \tilde{m} H.
\end{split}\label{eq:g-derivs}
\end{align}
The last expressions for both derivatives are derived using Eq.~\eqref{eq:m_tilde_a} and $Ht = 2/(3(1+\omega))$ for general $\omega$. The stationary point $a(\tau_s)$ is located at vanishing first derivative of the phase function: $g'(\tau_s) =0$. It leads to the implicit equation
\begin{align}\label{eq:position_SPA}
a_s = \frac{k}{(1 + t \partial_t )\tilde{m}j} \simeq \frac{n k}{2j \tilde m } \simeq \frac{n k}{2 j \tilde m_e} \lrfrac{a_s}{a_e}^{3w} \,,
\end{align}
where we defined the effective mass at the end of inflation $\tilde{m}_e \equiv \tilde{m} (a_e)$. We emphasize that $a_s$ in Eq.~\eqref{eq:position_SPA} is not identical to $a_k$ defined in Eq.~\eqref{eq:position_Bol} for general $n$. Only with $n=2$, they yield the same resonant criterion for graviton production: $k = a_s m_\phi$. We have used also that only positive $j$ can satisfy the stationary phase condition. Moreover, since $F_\text{rh}(\tau)$ is only present after inflation, we must impose $a_s \geq a_e$. Using Eq.~\eqref{eq:position_SPA}, we obtain the following constraints on the comoving momentum:
\begin{align}
    k 
    \begin{cases}
      > a_e m_\phi\, j , & n = 2 , \\[4pt]
      = \dfrac{\tilde a_e m_e j}{2} , & n = 4 , \\[6pt]
      < \dfrac{2j}{n}\, a_e \tilde m_e , & n > 4 .
    \end{cases}
\end{align}
In other words, the stationary point $\tau_s$ corresponds physically to the time when the physical momentum of the produced graviton, $k/a$, equals the inflaton effective mass.

With Eqs.~\eqref{eq:bogo-ana} and~\eqref{eq:g-derivs}, we can further simplify the absolute value of $\beta_k$ and omit the interference terms between different Fourier modes for now:
\begin{align}
    |\beta_k^\text{rh}| &\simeq \frac{\sqrt{\pi}}{4k\mpl^2} \sqrt{\frac{n(n+2)}{n-4}}  \sum_{j>0} \frac{ |V_j|}{\sqrt{j}} \left( \frac{a \rho_\phi}{\sqrt{\tilde{m} H}}  \right)_{a=a_s}, \notag \\
    &\simeq \frac{3\sqrt{\pi}}{8} \sqrt{\frac{n(n+2)}{n-4}}\sum_{j>0} |V_j| \lrfrac{H}{\tilde{m} j}^{3/2}, \notag \\  
    &\simeq \frac{3\sqrt{\pi}}{8} n \sqrt{\frac{n(n+2)}{|n-4|}}  \sum_{j>0} \frac{|V_j|}{j^{3/2}} \lrfrac{H_e}{\tilde{m}_e}^{3/2} \lrfrac{a_s}{a_e}^{-\frac{9}{n+2}}, \notag \\
    &\simeq \frac{3\sqrt{\pi}}{8} n \sqrt{\frac{n(n+2)}{|n-4|}}  \sum_{j>0} \frac{|V_j|}{j^{3/2}} \lrfrac{H_e}{\tilde{m}_e}^{3/2} \lrfrac{nk}{2j\tilde{m}_e a_e}^{\frac{9}{2 (n-4)}}.
    \label{eq:beta_k_app_j}
\end{align}
We have used $\rho_\phi \propto H^2 \propto a^{-3(1+\omega)}$ at second and third equalities, where a decay factor is omitted for $\rho_\phi$, because for general $n$, the decay cannot be described in terms of simple exponential \cite{Garcia:2020wiy}. Eq.~\eqref{eq:position_SPA} is used at second and fourth equalities and $\tilde{m} = \tilde{m}_e (a/a_e)^{-3\omega}$ is used at third equality. In the last step, the solution to Eq.~\eqref{eq:position_SPA} is inserted. We notice that just like in the Boltzmann case, the spectrum diverges with $n=4$.

There are similarities and differences between the Boltzmann method and stationary phase approximation within the Bogoliubov framework. Although they are based on different frameworks, the source for particle production is the same: the inflaton field described by $\rho_\phi$. The Feynman rules in Appendix \ref{sec:feynman-rules} and Eq.~\eqref{eq:a2R_der} both contain the energy density of the inflaton field and Fourier coefficients of its oscillation. Thus, for a given Fourier mode, they both lead to $|\beta_{k,j}|^2\propto |V_j|^2 \rho_\phi^2 / (\tilde{m} H)$, where the time-dependent quantities are evaluated at the characteristic scale factor $a=a_k$ or $a_s$, see Eq.~\eqref{eq:boltzmann-exact} and first line of Eq.~\eqref{eq:beta_k_app_j}. After applying the scaling of $\tilde{m}(a)$, we see that they both translate to the identical scaling in terms of $k$ for a given Fourier mode $ \sim k^{\frac{9}{2(n-4)}}$; see Eq.~\eqref{eq:boltzmann-power}  and Eq.~\eqref{eq:beta_k_app_j}. The characteristic scale factors are different in two methods: one is enforced by the Dirac delta for momentum conservation while the other one comes from the point of stationary phase. These two are in general not the same and they could eventually lead to different peak comoving momenta. Finally, we note that the stationary phase approximation in Eq.~\eqref{eq:bogo-ana} contains a phase factor and it leads to interferences of different Fourier modes, which is absent in our Boltzmann method in Eq.~\eqref{eq:boltzmann-exact}. 

For the simplest case with a quadratic potential ($n=2$), the effective oscillation frequency is essentially constant, and
these two methods deliver identical spectra. In this case, the dominant contribution arises from the $j=1$ mode with $V_1 = -1/4$. As a result, Eq.~\eqref{eq:beta_k_app_j} reduces to 
\begin{equation}\label{eq:beta_k_app}
    |\beta_k^{\text{rh}}| \simeq \frac{3\sqrt{\pi}}{8} \left(\frac{H_e}{m_\phi}\right)^{3/2}\left(\frac{a_e m_\phi}{k}\right)^{9/4} \exp \left(-\frac{\Gamma_\phi}{H_e} \lrfrac{k}{a_em_\phi}^{3/2} \right) \,,
\end{equation}
which is in agreement with Eq.~\eqref{eq:boltzmann-ana-2} from Boltzmann approach. The exponential factor is reintroduced compared to Eq.~\eqref{eq:beta_k_app_j}. We note the scaling with $k$ and the dependence on the model parameters in Eq.~\eqref{eq:beta_k_app} are consistent with Eq.~(51) of Ref.~\cite{Kaneta:2022gug} for the case $n=2$. 
 
Several comments are in order before moving on.
% Although the curvature term $a^{2}R/6$ is retained in the mode frequency appearing in Eq.~\eqref{eq:beta_k_n}, 
The stationary-phase approximation isolates only those contributions arising from
rapid oscillations of the background that generate stationary points of the phase. As a result, it discards graviton production sourced by the slow, background-driven evolution of the expansion as well as the production coming from transition from inflation to reheating with the equation of state parameter $\omega$. Depending on the specific background dynamics, these contributions can be important. These effects will be discussed in the following section. 

\subsection{Solution for Bogoliubov Coefficient: Transition Contribution}
\label{sec:transition}

In this subsection we investigate the transition contribution, aiming for deriving an analytical solution to Eq.~\eqref{eq:beta_Fprime_app} for
graviton production associated with the transition from inflation to reheating. The corresponding Bogoliubov coefficients can be written as 
\begin{equation}
\beta_k^{\rm tr} \simeq - \frac{1}{4k^2} \int^{\tau_f}_{\tau_i} d\tau \, F'_{\rm tr}(\tau)e^{-2ik\tau}.
\label{eq:beta_transition_def_app}
\end{equation}

\begin{figure}
    \centering
    \includegraphics[width=0.7\linewidth]{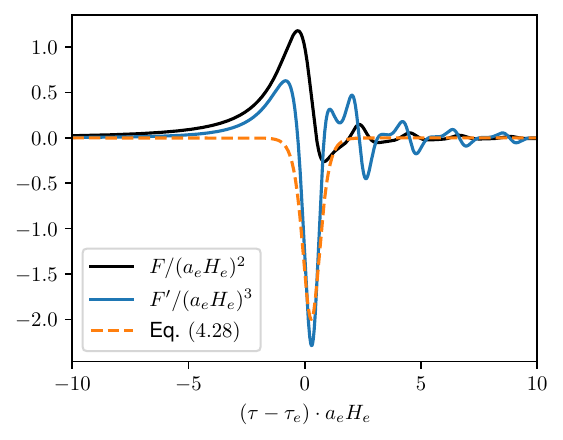}
    \caption{Evolution of the function $F(\tau) =a^2 R/6$ (solid black), its derivative $F'(\tau)$ with respect to  time (solid blue), and the fit of $F'(\tau)$ (dashed orange) using Eq.~\eqref{eq:Ftr_sech_app}. The curves are drawn with $n=4$ T model.}
    \label{fig:F_fit}
\end{figure}

The transition from inflation to reheating is localized in time and produces a
brief but significant violation of adiabaticity near $\tau=\tau_e$. As shown  by the solid black curve in Fig.~\ref{fig:F_fit}, the quantity $F(\tau)=a^2R/6$ develops a localized and highly asymmetrical peak
around $\tau_e$ before relaxing toward the oscillatory reheating regime. Its derivative $F'(\tau)$ (the orange dashed line) can be well described by the function $\mathrm{sech}^2=\cosh^{-2}$ around $\tau_e$, where it has a sharp trough and the non-adiabatic effects are most pronounced. Motivated by this observation, we model the transition contribution using the following ansatz

\begin{equation}
\label{eq:Ftr_sech_app}
F^{\prime}_{\rm tr}(\tau)=
A_{\rm pk}\,
\mathrm{sech}^2\left(\frac{\tau-\tau_e}{\Delta\tau}\right),
\end{equation}
where $\Delta\tau$ characterizes the duration of the transition and $A_{\rm pk}$ denotes its amplitude (will be determined below). In Fig.~\ref{fig:F_fit}, we show a fit with $\Delta \tau = 0.5\, a_e H_e$ for $n=4$ T model, which provides a good description of the behavior of $F'(\tau)$ around $\tau_e$. We want to emphasize that the complete $F(\tau)$ and $F'(\tau)$ are shown in Fig.~\ref{fig:F_fit}, which contain both the transition and oscillation contributions. In this section, we aim to capture the transition contributions, manifested as the trough in $F'(\tau)$.
% the small peak in $F'(\tau)$ just before $\tau_e$ is not considered, but should be subdominant and the oscillations after $\tau_e$ are considered in the previous subsection already. 
We have also verified that the same value of $\Delta \tau$ describes the $n=6$ case equally well. Physically, this ansatz captures the fact that particle production from background transition is controlled by a short, localized burst of non-adiabatic evolution centered at the end of inflation. 

Introducing the dimensionless variable
$x=(\tau-\tau_e)/\Delta\tau$, and substituting Eq.~\eqref{eq:Ftr_sech_app} into
Eq.~\eqref{eq:beta_transition_def_app}, we obtain
\begin{equation}
\beta_k^{\rm tr}\simeq
\frac{-iA_{\rm pk}\Delta\tau}{4k^2}
e^{-2ik\tau_e}
\int_{-\infty}^{\infty}dx\,
\mathrm{sech}^2x\,e^{-2ik\Delta\tau x}.
\end{equation}
Utilizing the standard Fourier integral $\int_{-\infty}^{\infty}dx\,\mathrm{sech}^2x\,e^{-iqx}= \pi q/\sinh(\pi q/2)$,
with $q=2k\Delta\tau$, the Bogoliubov coefficient becomes
\begin{equation}
\label{eq:beta_transition_final_new}
\beta_k^{\rm tr}
\simeq
\frac{-i\pi A_{\rm pk}(\Delta\tau)^2}{2k}
\frac{e^{-2ik\tau_e}}{\sinh(\pi k\Delta\tau)}.
\end{equation}
Eq.~\eqref{eq:beta_transition_final_new} has a transparent interpretation. Since $\beta_k^{\rm tr}$ is essentially the Fourier transform  (Eq.~\eqref{eq:beta_transition_def_app}) of the localized burst $F^{\prime}_{\rm tr}(\tau)$ of length $\Delta \tau$, its magnitude is controlled by how rapidly the phase $e^{-2ik\tau}$ varies across the transition. If $k\Delta\tau\ll1$, the phase
changes little over the interval $\Delta\tau$ and the contributions add
constructively, leading to efficient  particle production. By contrast, if
$k\Delta\tau\gg1$ the phase oscillates rapidly during the burst, producing strong
cancellations and therefore exponential suppression.

The magnitude of the Bogoliubov coefficients therefore reads 
\begin{equation}
\label{eq:beta_transition_mag_new}
|\beta_k^{\rm tr}|
\simeq
\frac{\pi |A_{\rm pk}|(\Delta\tau)^2}{2k}
\frac{1}{\sinh(\pi k\Delta\tau)}\,,
\end{equation}
which exhibits the following limiting behavior,
\begin{equation}
|\beta_k^{\rm tr}|
\simeq |A_{\rm pk}|
\begin{cases}
\dfrac{\Delta\tau}{2k^2}, & k\Delta\tau\ll1, \\ 
\frac{\pi (\Delta\tau)^2}{k} e^{-\pi k\Delta\tau}, & k\Delta\tau\gg1.
\end{cases}
\label{eq:beta_transition_limits}
\end{equation}

Our treatment of $F'_\text{tr}(\tau)$ in Eq.~\eqref{eq:Ftr_sech_app} corresponds to fitting $F(\tau)$ with a $\tanh$-function. The fit of $F(\tau)$ thus takes constant values before and after the transition. We now determine the amplitude parameter $A_{\rm pk}$ by computing the integral\footnote{The infinite integral limit should be understood as an approximation of finite limit. It gives almost identical results when the integral limit is taken to be $\mathcal{O}(1) \Delta \tau$, with the advantage that integral yields an integer. }
\begin{equation}
\int_{-\infty}^{+\infty} d\tau\,F'_{\rm tr}(\tau)
=
A_{\rm pk}\,\Delta\tau
\int_{-\infty}^{+\infty} dx\,\sech^2 x
= 2 A_{\text{pk}}\Delta \tau\,.
\label{eq:Ftr_integral}
\end{equation}
On the other hand, by construction the same integral must equal the total jump of $F$ across the transition,
\begin{equation}
\int_{-\infty}^{+\infty} d\tau\,F'_{\rm tr}(\tau)
=
F_+ - F_-,
\label{eq:F_jump_match}
\end{equation}
where $F_-$ denotes the inflationary-side value and $F_+$ the value just after the transition into the oscillatory reheating regime. Matching Eqs.~\eqref{eq:Ftr_integral} and \eqref{eq:F_jump_match}, we obtain the general result
\begin{equation}
A_{\rm pk}
=
\frac{F_+ - F_-}{2 \Delta\tau}.
\label{eq:Apk_general_match}
\end{equation}

We now evaluate $F_\pm$ for a potential $V(\phi)\propto \phi^n$ using $F\simeq a^2 H^2 (1-3\omega)/2$. 
During inflation one has $w\simeq -1$, and therefore
\begin{equation}
F_-
\simeq
\frac{a_e^2 H_e^2}{2}\bigl(1-3(-1)\bigr)
=
2a_e^2H_e^2.
\label{eq:Fminus}
\end{equation}
After the transition, the oscillating inflaton condensate behaves as a fluid with averaged equation-of-state parameter $w =(n-2)/(n+2)$
which yields
\begin{equation}
F_+
\simeq
\frac{a_e^2 H_e^2}{2}
\left(1-3\frac{n-2}{n+2}\right)
=
a_e^2 H_e^2\,\frac{4-n}{n+2}.
\label{eq:Fplus}
\end{equation}
Here we used the fact that the transition is short, so that $a^2H^2$ can be approximated by its value at $\tau=\tau_e$ around the burst.

Combining Eqs.~\eqref{eq:Fminus} and \eqref{eq:Fplus}, the net change is
\begin{equation}
F_+ - F_-
=
a_e^2H_e^2\left(\frac{4-n}{n+2}-2\right)
=
-\,a_e^2H_e^2\,\frac{3n}{n+2}.
\label{eq:Fjump_final}
\end{equation}
Substituting this into Eq.~\eqref{eq:Apk_general_match}, we finally obtain
\begin{equation}
A_{\rm pk}
\simeq
-\,\frac{3n}{2 (n+2)}
\frac{a_e^2H_e^2}{\Delta\tau},
\label{eq:Apk_final_match}
\end{equation}
which also has a transparent interpretation: $A_{\rm pk}$ is fixed by the total drop of $F(\tau)$ during the transition divided by $\Delta\tau$. 
% In particular, a sharper transition (i.e. smaller $\Delta\tau$) corresponds to a larger $|A_{\rm pk}|$, and the dependence on $n$ reflects the change in the effective post-inflationary equation of state. This analytical estimate is used in Fig.~\ref{fig:F_fit}. The drop in $F(\tau)$ is visible from Fig.~\ref{fig:F_fit}, where it has peak value $\sim (a_e H_e)^2$ and drops to more or less zero as shown by the black line. Exactly this drop translates into a dip in the derivative $F'(\tau)$, which we model in this section. \cw{Although the estimate in Eq.~\eqref{eq:Fminus} doesn't agree with Fig.~\ref{fig:F_fit} perfectly, the eventual height $A_\text{pk}$ fits the $F'(\tau)$ well.} 
In particular, a sharper transition, corresponding to a smaller $\Delta\tau$, leads to a larger value of $|A_{\rm pk}|$, while the $n$ dependence encodes the change in the effective post-inflationary equation of state. This analytical estimate is used in Fig.~\ref{fig:F_fit}. As seen there, $F(\tau)$ reaches a value of order $(a_e H_e)^2$ around the end of inflation and then drops rapidly toward a value close to zero, as indicated by the black curve. It is precisely this drop that gives rise to the localized dip in $F'(\tau)$, which we model in this section. Although the estimate in Eq.~\eqref{eq:Fminus} does not reproduce Fig.~\ref{fig:F_fit} exactly, it captures the relevant overall change in $F(\tau)$ and, in turn, yields an accurate estimate for the height $A_{\rm pk}$ that governs the profile of $F'(\tau)$.

Using Eq.~\eqref{eq:Apk_final_match}, we can rewrite Eq.~\eqref{eq:beta_transition_mag_new} as
\begin{equation}
\label{eq:beta_transition_final}
|\beta_k^{\rm tr}|
\simeq  
\frac{3n}{4(n+2)}
\left(\frac{a_e H_e}{k}\right)
\frac{\Delta\tau \, a_e H_e \pi}{\sinh(\pi k\Delta\tau)}\,.
\end{equation}
 Here, $\Delta \tau \sim \order{1}/ (a_e H_e)$ generally, which will be specified in next section when we present the numerical spectrum.

Eq.~\eqref{eq:beta_transition_final} is the central result of this subsection. To the best of our knowledge, such a compact parametric expression has not appeared in the literature. We note that $|\beta_k^{\rm tr}|$ depends only weakly on $n$ in the regime where the transition contributions are prominent. This reflects the fact that the transition occurs while the inflaton is still far from the minimum of its potential. On the other hand, the oscillatory contribution is more sensitive to the value of $n$, and its overall amplitude decreases as $n$ increases. This can be understood from the structure of the Fourier decomposition of the inflaton oscillation. For $n=2$, graviton production is dominated by the first Fourier mode, which efficiently sources modes around $k \sim a_e m_\phi$. For $n>2$, however, the anharmonic oscillations generate a broader spectrum of Fourier modes, and producing gravitons at comparable momenta requires contributions from higher harmonics. This leads to a reduced overall efficiency of the oscillatory production mechanism suppressed by the Fourier coefficients $V_j$  (cf.~Eq.~\eqref{eq:boltzmann-power}). Consequently, the transition contributions becomes dominant with increasing $n$. This will be demonstrated more explicitly in the numerical results presented in the next section, as well as in Appendix~\ref{app:dominance}.

 Before closing this section, we briefly compare our treatment with existing approaches in the literature. In particular, Ref.~\cite{deGarciaMaia:1993ck} employed a sudden matching approximation. By contrast, our parametrization in Eq.~\eqref{eq:Ftr_sech_app} goes beyond this limit by incorporating the effects of finite transition duration, thereby capturing the gradual breakdown of adiabaticity. Ref.~\cite{Pi:2024kpw} adopted a different strategy, in which the background evolution is modeled by a smooth analytical fit of $aH$ across the transition, with a parametrized equation-of-state $\omega$. In our approach, we directly treat the localized source of particle production by modeling the burst in $F'(\tau)$ (i.e. time derivative of the background), as in Eq.~\eqref{eq:Ftr_sech_app}. Being different from Ref.~\cite{Pi:2024kpw}, where the Bogoliubov coefficients are obtained by solving the mode equation~\eqref{eq:EOM-h} in a smoothed background, we instead adopt a complementary approach. In our case, the coefficients are computed by evaluating the integral in Eq.~\eqref{eq:beta_transition_def_app} using an approximate form of $F'(\tau)$, thereby avoiding solving the full mode equation Eq.~\eqref{eq:EOM-h} explicitly.

As we will show in the next section, the analytical approximation in Eq.~\eqref{eq:beta_transition_final}, constructed to capture the localized variation of $F'(\tau)$ associated with the finite-time breakdown of adiabaticity, provides a description consistent with the numerical Bogoliubov coefficients over the relevant range. 

%%%%%%%%%%%%%%%%%%%%%%%%%%%
\section{Comparison between Boltzmann and Bogoliubov}\label{sec:comparison}
%%%%%%%%%%%%%%%%%%%%%%%%%%%

Before quantitatively  comparing the two formalisms, we shall make it explicit that the quantities $f_h$ in Section~\ref{sec:boltz} and $|\beta_k|^2$ in Section~\ref{sec:bogo} have the same physical meaning, despite different definitions of the graviton field. The phase space distribution $f_h(a,k)$ is defined such that the total comoving energy density carried by gravitational waves is given by \cite{Xu:2025wjq}
\begin{equation} \label{eq:rhoGW-int}
    a^4 \rho_h(a) = g_h \int \frac{\dd[3]{k}}{(2\pi)^3}\, k f_h(a, k)\,.
\end{equation}
Here, $g_h =2$ denotes the two degrees of freedom for massless gravitons, and $k$ corresponds the comoving momentum. On the other hand, the GW energy density in the Bogoliubov approach is more subtle. As usual, the energy density is defined as the $(0,0)$-component of the energy--momentum tensor. Since we are dealing with a quantized field, we evaluate the vacuum expectation value of the energy--momentum tensor with respect to the cosmic rest frame. As in Minkowski quantum field theory for a scalar field, ultraviolet divergences are present and must be renormalized. After renormalization by normal ordering, the comoving energy density of gravitational waves is given by~\cite{Kolb:2023ydq}
\begin{align}\label{eq:rho_h}
a^4\,\rho_h(\tau) = g_h \int \frac{d^3k}{(2\pi)^3} \omega_k|\beta_k (\tau)|^2\,.
\end{align}
For modes well inside the horizon, we see that $\omega_k \to k$, and consequently $|\beta_k|^2 \to f_h(k)$. As a result, we see that $f_h(a,k)$ and $|\beta_k|^2$ may carry similar physical information. Note that the tachyonic instability due to negative $a^2R/6$ is only temporary and it doesn't affect the physical interpretation here. In the previous sections, we compared these two quantities at a qualitative level; we proceed to a quantitative comparison in this section. 

For the numerical calculations, we must specify the background evolution explicitly. As indicated in Eq.~\eqref{eq:background}, we include the inflaton decay rate $\Gamma_\phi$ in the background inflaton equation of motion to account for inflaton decay. It is therefore necessary to provide an explicit expression for $\Gamma_\phi$.  We will adopt a fermionic decay channel as our benchmark, although we expect similar results with other channels.  For potential well approximated by quadratic term near the origin, the decay width of inflaton to fermions $\phi \rightarrow \bar{f} f$ via a Yukawa coupling $y$ simply reads $  \Gamma_\phi \simeq y^2 m_\phi /8\pi,$ where we have omitted the mass of the daughter fields. For a generalized potential with $n > 2$ in Eq.~\eqref{eq:vphi}, the situation becomes more complicated. First the inflaton effective mass $d^2 V/{d\phi}^2$ depends on the oscillation amplitude, which decreases with the cosmic expansion. Second, there exists an additional correction due to the inflaton anharmonic oscillation. Here, we follow Ref.~\cite{Garcia:2020wiy}, in which the decay width via the same Yukawa coupling can be written as:
\begin{equation}
    \Gamma_\phi (t) = \sqrt{n(n-1)} \lambda^{\frac{1}{n}}  \frac{y_\text{eff}^2}{8\pi} \lrfrac{\rho_\phi (t)}{\mpl^4}^{\frac{1}{2}-\frac{1}{n}} \mpl,
    \label{eq:generalized-potential-Gamma}
\end{equation}
where $y_\text{eff}$ is the effective Yukawa coupling which includes corrections from inflaton anharmonic oscillations and can only be computed numerically; see Refs.~\cite{Ichikawa:2008ne, Garcia:2020wiy} for more details. $\lambda$ is the coefficients in the potential in Eq.~\eqref{eq:vphi}. The reheating temperature is given by \cite{Garcia:2020wiy}
\begin{equation}
    T_\text{rh} = \lrfrac{30}{g_* \pi^2}^{\frac{1}{4}} \left[ \frac{n\sqrt{3n(n-1)}}{7-n} \lambda^{\frac{1}{n}} \frac{y_\text{eff}^2}{8\pi} \right]^{\frac{n}{4}} \mpl,
    \label{eq:generalized-potential-Trh}
\end{equation}
for $ n < 7$. Combining Eqs.~\eqref{eq:generalized-potential-Gamma} and \eqref{eq:generalized-potential-Trh},  we can express the decay rate in terms of the reheating temperature
\begin{equation}
    \Gamma_\phi (t) = \frac{7-n}{\sqrt{3}n} \lrfrac{T_\text{rh}}{\mpl}^{\frac{4}{n}} \lrfrac{30}{g_* \pi^2}^{-\frac{1}{n}} \lrfrac{\rho_\phi(t)}{\mpl^4}^{\frac{1}{2}-\frac{1}{n}} \mpl,
    \label{eq:generalized-potential-Gamma-Trh}
\end{equation}
which will be used for Eq.~\eqref{eq:background}. Some details on the numerical simulation are summarized in the Appendix~\ref{app:num}. Moreover, the Boltzmann formula in Eq.~\eqref{eq:boltzmann-exact}, and the Bogoliubov expression in Eq.~\eqref{eq:bogo-exact} are largely agnostic about the underlying inflationary model. For concreteness, we adopt a well-motivated scenario, namely the $T$-model $\alpha$-attractor for illustration purpose,  and model-specific details  are summarized in Appendix~\ref{app:T_model}.

%%%%%%%%%%%%%%%%%%%%%%%%%%%%%%%%%%%%%%%%%%%%%%%%%%%%%%%%%%%%%%%%%%%%
\subsection{$n=2$}
%%%%%%%%%%%%%%%%%%%%%%%%%%%%%%%%%%%%%%%%%%%%%%%%%%%%%%%%%%%%%%%%%%%%

In this subsection, we investigate the scenario with $n=2$, focusing on the quantitative comparison between the gravitational-wave spectrum $f_h$ obtained from the Boltzmann approach and $|\beta_k|^2$ computed using the Bogoliubov method. 
\begin{figure}[ht!]
    \centering
\includegraphics[width=0.7\linewidth]{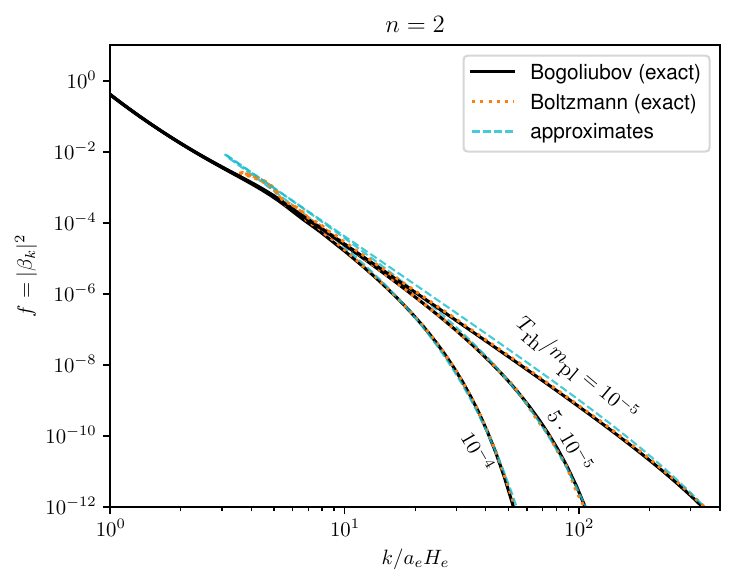}\\
    \caption{Phase space spectra with $n=2$. Exact Bogoliubov refers to the spectrum computed from numerically solving Eq.~\eqref{eq:EOM-h}. Exact Boltzmann spectrum is obtained from Eq.~\eqref{eq:exact_boltzmann} using the Fourier coefficients. Approximates correspond to Eq.~\eqref{eq:boltzmann-ana-2} in the Boltzmann formalism or equivalently Eq.~\eqref{eq:beta_k_app} in the Bogoliubov method. }
    \label{fig:gw-spec-quadratic} 
\end{figure}

The graviton spectra obtained using the Boltzmann and Bogoliubov methods are shown in Fig.~\ref{fig:gw-spec-quadratic}. We take the inflaton mass to be $m_\phi\simeq\num{4.6e-6} \mpl$, corresponding to the $T$-model prediction of a tensor-to-scalar ratio $r=0.01$. From left to right, the panels show reheating temperatures $\Trh=10^{-4}\mpl$, $5\times10^{-5}\mpl$, and $10^{-5}\mpl$.

The black solid curve is obtained using the Bogoliubov method by numerically solving Eq.~\eqref{eq:EOM-h}. The orange dotted curve shows the exact Boltzmann spectrum computed numerically from the Fourier coefficients, Eq.~\eqref{eq:exact_boltzmann}. The cyan dashed curve shows the analytic approximation, Eq.~\eqref{eq:boltzmann-ana-2} in the Boltzmann formalism or equivalently Eq.~\eqref{eq:beta_k_app} in the Bogoliubov approach using the stationary point approximation.

The spectra in Fig.~\ref{fig:gw-spec-quadratic} follow a scaling $k^{-9/2}$ until the exponential decay due to the inflaton condensate decay start to become effective. This behavior is consistent with the analytic results in Eqs.~\eqref{eq:boltzmann-ana-2} and~\eqref{eq:beta_k_app}. At late times, the spectrum is exponentially suppressed, reflecting the decay of the inflaton condensate. 

The Boltzmann and Bogoliubov results agree closely at large $k$. As discussed in the Section \ref{sec:transition} and Appendix \ref{app:dominance}, the transition contribution is negligible compared to oscillation contribution for $n=2$, explaining why the stationary-phase approximation accurately reproduces the full Bogoliubov result. Since the stationary-phase approximation corresponds to the Boltzmann treatment in this regime, this also accounts for the agreement between the two approaches.  As shown in Fig.~\ref{fig:gw-spec-quadratic}, these conclusions are robust and remain unchanged under variations of the reheating temperature. Finally, we also find non-perturbative structure in $|\beta_k|^2$ for long-wavelength modes with $k\ll a_eH_e$, associated with the tachyonic effective frequency $\omega_k^2$ defined in Eq.~\eqref{eq:effective_frequency}, although this feature is not visible in the figure.

Having established that $f_h(k)=|\beta_k|^2$ in both approaches at large $k$ for $n=2$, we now turn to steeper potentials.

%%%%%%%%%%%%%%%%%%%%%%%%%%%%%%%%%%%%%%%%%%%%%%%%%%%%%%%%%%%%%%%%%%%
\subsection{$n=4$}
%%%%%%%%%%%%%%%%%%%%%%%%%%%%%%%%%%%%%%%%%%%%%%%%%%%%%%%%%%%%%%%%%%

\begin{figure}
    \centering
    \includegraphics[width=0.7\linewidth]{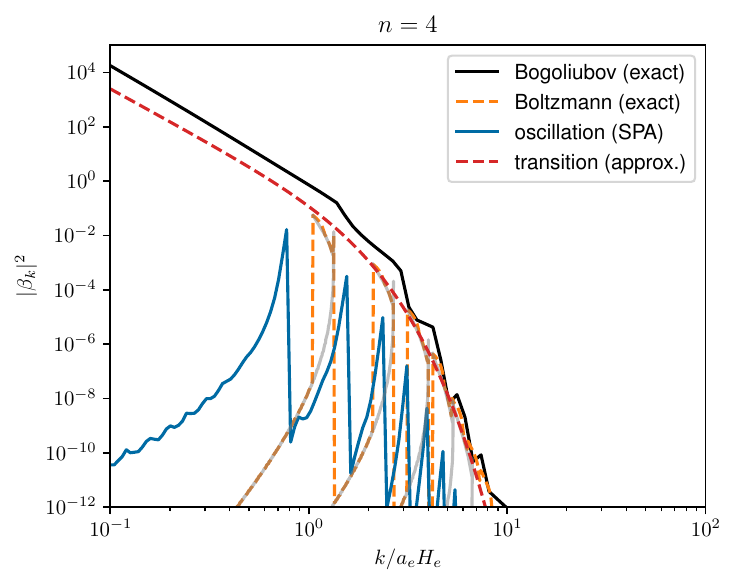}\\%
    \caption{Graviton phase-space spectrum for $n=4$ with $\Trh=10^{-5}\mpl$. The black curve shows the full Bogoliubov result from Eq.~\eqref{eq:bogo-exact}. The dashed orange curve shows the sum over Fourier modes from Eq.~\eqref{eq:boltzmann-exact}, while the gray curves show the individual mode contributions. The blue curve shows the stationary-phase approximation (SPA) from Eq.~\eqref{eq:bogo-ana}. The red dashed curve shows the transition contribution from Eq.~\eqref{eq:beta_transition_final}.
    }
    \label{fig:quartic-spec}
\end{figure}

In this subsection, we investigate the scenario with $n=4$. The spectra obtained from the Boltzmann and Bogoliubov methods are shown in Fig.~\ref{fig:quartic-spec}. As in the $n=2$ case, the tensor-to-scalar ratio is taken to be  $r=0.01$; the reheating temperature is fixed to be $\Trh = 10^{-5} \mpl$. 

The black solid curve shows the full numerical result obtained in the Bogoliubov formalism by solving Eq.~\eqref{eq:EOM-h} using Eq.~\eqref{eq:bogo-exact}. The dashed orange curve represents the sum over all Fourier modes from Eq.~\eqref{eq:boltzmann-exact}, while the gray curves show the individual Fourier mode contributions. The solid blue curve corresponds to the approximate spectrum from Eq.~\eqref{eq:bogo-ana} evaluated using the stationary-phase approximation (SPA). The red dashed curve denotes the transition contribution based on Eq.~\eqref{eq:beta_transition_final}. Compared to the previous $n=2$ case, several new features emerge as we will explain below.

To begin with, the Boltzmann result no longer aligns with the SPA result, and the characteristic momenta of produced gravitons are shifted toward larger values in the Boltzmann spectrum. This can be seen from our analytic estimates in Eqs.~\eqref{eq:position_Bol} and~\eqref{eq:position_SPA}. Since the time derivative of the effective inflaton mass is not included in the determination of comoving momentum in the Boltzmann approach, the two methods select different production times and therefore different characteristic momenta. Apart from this momentum shift and the associated change in normalization, however, both approaches exhibit identical scaling behavior with $ k^{9/[2(n-4)]}$ for a single Fourier mode for small $k$ (cf. Eq.~\eqref{eq:boltzmann-power} and Eq.~\eqref{eq:beta_k_app_j}).

For each Fourier harmonic $j$, the Boltzmann spectrum follows a power law extending toward lower momenta, as shown by the gray dashed line. In the analytic quartic solution one has $\tilde m\propto a^{-1}$, so that $a\tilde m$ would be constant and many oscillations would contribute to the exactly same comoving momentum. According to Eq.~\eqref{eq:boltzmann-power}, this would lead to a formally divergent spectrum consisting of multiple sharp peaks. In practice, however, we find numerically that the oscillation frequency behaves as $\tilde m\sim a^{-0.89}$ initially during reheating, so $a\tilde m$ is not exactly constant. As a result, the overlap between contributions from different oscillations is reduced and the peaks are smoothed, yielding a finite enhancement $C(k\simeq a_eH_e)\sim\order{10}$ in the Boltzmann result. Another important feature is that graviton production as predicted by Boltzmann method is dominated by the earliest oscillations, when the inflaton energy density is largest and decay is still negligible. As a result, the spectral shape is largely determined by the first few oscillations and depends only weakly on the decay width $\Gamma_\phi$. 

Finally, and most importantly, as clearly seen in Fig.~\ref{fig:quartic-spec}, the Boltzmann result deviates significantly from the full Bogoliubov spectrum, and the stationary-phase approximation fails as well. As discussed in Section \ref{sec:transition} and Appendix \ref{app:dominance}, this reflects the dominance of the non-oscillatory transition contribution captured only by the full Bogoliubov treatment. In the momentum range of interest, graviton production is therefore controlled primarily by the inflation–reheating transition (see Sec.~\ref{sec:transition}). This can be seen in Fig.~\ref{fig:quartic-spec}, where the red dashed curve drawn using Eq.~\eqref{eq:beta_transition_final} with $\Delta\tau = 0.5 \, a_e H_e$ closely tracks the full Bogoliubov result for $k\gtrsim a_eH_e$. For $k\lesssim a_eH_e$, additional effect such as non-perturbative contributions relevant, explaining why the transition component is subdominant in that regime. For this exact reason, we only show the spectra with a single $\Trh$. Because the production is primarily from the transition, the inflaton decay width $\Gamma_\phi$ has barely any effects on the graviton spectrum. We also remind the reader that the red dashed line (Eq.~\eqref{eq:beta_transition_final}) was derived under the assumption $|\beta_k| \ll 1$. It is therefore expected to agree with the full numerical results only within its regime of validity, and to deviate once this condition is violated.

Before closing this subsection, we note that the numerical Bogoliubov spectrum in Fig.~\ref{fig:quartic-spec} scales as $k^{-4}$ in the low momentum range. Consequently, the integral in Eq.~\eqref{eq:rhoGW-int} or equivalently~\eqref{eq:rho_h} would be IR-divergent. This can be regulated by imposing the IR cutoff $k\geq k_\text{IR}=a_0 H_0$, corresponding to the current comoving horizon size \cite{Kolb:2023ydq, Garcia:2023awt}.

We now turn to the case $n=6$, where the discrepancy between the Boltzmann and Bogoliubov results becomes even more pronounced.

%%%%%%%%%%%%%%%%%%%%%%%%%%%%%%%%%%%%%%%%%%%%%%%%%%%%%%%%%%%%%%%%%%%%
\subsection{$n=6$}
%%%%%%%%%%%%%%%%%%%%%%%%%%%%%%%%%%%%%%%%%%%%%%%%%%%%%%%%%%%%%%%%%%%%

\begin{figure}[ht!]
    \centering
    \includegraphics[width=0.7\linewidth]{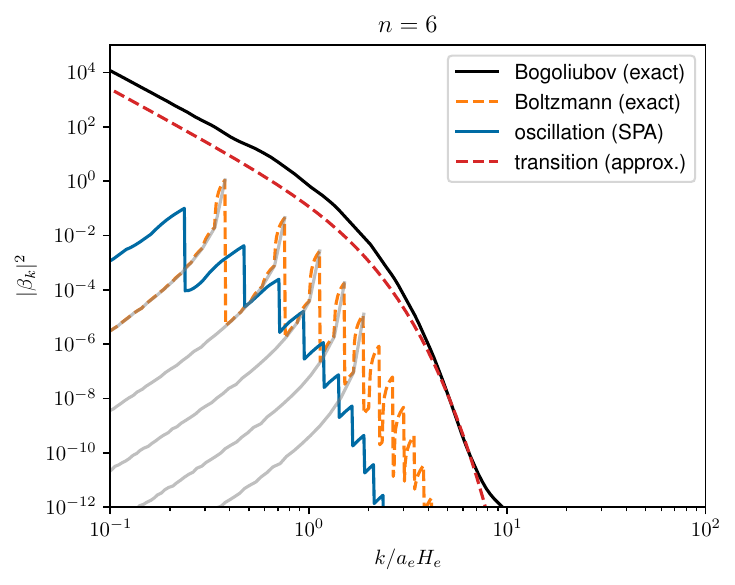}%
    \caption{Graviton phase space spectra for $n=6$. The curves have the same meaning as in Fig.~\ref{fig:quartic-spec}. }
    \label{fig:sextic-spec}
\end{figure}

In this subsection we consider the case $n=6$. The spectra are shown in Fig.~\ref{fig:sextic-spec}, with line styles identical to those in Fig.~\ref{fig:quartic-spec}. As before, we take $r=0.01$ and $\Trh=10^{-5}\mpl$.

Compared to the $n=4$ case, the hierarchy between the different contributions becomes even more pronounced. Over essentially the entire momentum range shown, the Boltzmann result deviates significantly from the full Bogoliubov spectrum because the non-oscillatory transition contribution dominates graviton production. For $n=6$, this dominance extends over a wider momentum range than in the quartic case. This can be seen in Fig.~\ref{fig:sextic-spec}, where the red dashed curve closely tracks the full Bogoliubov result across most of the displayed range. 

As in the $n=4$ case, the Boltzmann result also does not fully align with the SPA.  The characteristic momenta are shifted toward larger values in the Boltzmann spectrum as shown by the orange curve. The overall envelope of both results remains determined by the Fourier coefficients, and the predicted power-law behavior extends only toward low momenta shown by the gray line. We note that the $n=6$ case predicts the equation of state $\omega=1/2$, hence there is no divergence feature nor amplification in the spectrum, making the Boltzmann spectrum lower than that in the $n=4$ case. Moreover, the full Bogoliubov result (black curve) becomes noticeably smoother for $n=6$ in Fig.~\ref{fig:sextic-spec} compared to the $n=4$ case in Fig.~\ref{fig:quartic-spec}. This can be understood from the relative importance of the two contributions discussed in Sec.~\ref{sec:transition}. As $n$ increases, the oscillatory contribution becomes less important, and the spectrum is increasingly controlled by the transition contribution. Since the latter arises from a single localized non-adiabatic event, it leads to a smooth momentum dependence. Consequently, the full Bogoliubov spectrum becomes smoother for larger $n$.

Overall, the discussion above reinforces the conclusion that for $n>2$ the Boltzmann description fails to reproduce the full Bogoliubov spectrum. In addition, the stationary-phase approximation becomes unreliable for describing the full particle production spectrum once the transition contribution dominates.

\section{Gravitational Wave Spectrum}\label{sec:gw_sp}
Having established that only the full Bogoliubov method provides an accurate description of the phase space distributions, we now turn to the resulting gravitational-wave (GW) signals derived from the Bogoliubov coefficients. We begin with the definition of the GW energy density spectrum,
\begin{align}\label{eq:OGW_def}
    \ogw(k) h^2 \equiv \frac{1}{\rho_c}\,\frac{d\rho_h}{d\ln k}
    = \frac{h^2}{\rho_c}\frac{\rho_{h, k}}{a^4}\,,
\end{align}
where $\rho_c = 3\mpl^2 H^2$ denotes the critical energy density, and $\rho_{h,k}$ is defined as
\begin{equation}
    \rho_{h, k} \equiv a^4 \dv{\rho_h}{\ln k}.
\end{equation}
The modes of interest are well inside the horizon at the end of reheating, and the physical graviton energy density therefore redshifts just as radiation once produced. Consequently, $\rho_{h,k}$ remains constant thereafter, and the present-day spectrum can then be written as
\begin{align}\label{eq:OGW_a0}
    \Omega_{\rm GW,0} h^2
    &= \frac{h^2}{\rho_{c,0}}\frac{\rho_{h, k}(a_0)}{a_0^4}= \frac{\rho_{h,k}}{a_e^4 H_e^4}
    \left(\frac{a_e}{a_{\rm rh}} \right)^4
    \left(\frac{a_{\rm rh}}{a_0} \right)^4
    \frac{h^2}{\rho_{c,0}} H_e^4 \notag \\
    &\simeq \num{1.71e-5} \cdot
    \left(\frac{\rho_{h, k}}{a_e^4 H_e^4}\right)
    \left(\frac{a_e}{a_{\rm rh}}\right)^4
    \left(\frac{H_e}{T_{\rm rh}}\right)^4\,,
\end{align}
where we have used the entropy conservation after reheating with
\begin{equation}\label{eq:entropy-scale}
\frac{a_{\rm rh}}{a_0}
=
\left(\frac{g_{*s,0}}{g_{*s,{\rm rh}}}\right)^{1/3}
\frac{T_0}{T_{\rm rh}}\,.
\end{equation}
For numerical factors, we take $T_0\simeq\SI{2.35e-4}{\eV}$, $g_{*s,0}\simeq 3.91$, $g_{*s,{\rm rh}}=106.75$, and $\rho_{c,0}/h^2 \simeq \SI{1.78e-46}{GeV\tothe{4}}$~\cite{ParticleDataGroup:2024cfk}.

The GW frequency at present is
\begin{align}
f
&= \frac{k}{2\pi a_0}
= \frac{1}{2\pi}\frac{k}{a_e H_e}
\frac{a_e}{a_{\rm rh}}
\frac{a_{\rm rh}}{a_0}
H_e \notag \\
&\simeq \SI{18.86}{\giga\hertz}\times
\left(\frac{k}{a_e H_e}\right)
\left(\frac{a_e}{a_{\rm rh}}\right)
\left(\frac{H_e}{T_{\rm rh}}\right),
\label{eq:f-clean}
\end{align}
where Eq.~\eqref{eq:entropy-scale} is used.

The numerical simulations to obtain $\beta_k$ are terminated at $\Omega_\phi \equiv \rho_\phi/(3 \mpl^2 H^2)=10^{-3}$, by which time the inflaton energy density has decayed exponentially and GW production has effectively converged, with $\rho_{h,k}$ approaching a constant. We use this asymptotic value of $\rho_{h,k}$ to evaluate the GW spectrum at present in Eq.~\eqref{eq:OGW_a0}.

\begin{figure}
    \centering
    \includegraphics[width=0.7\linewidth]{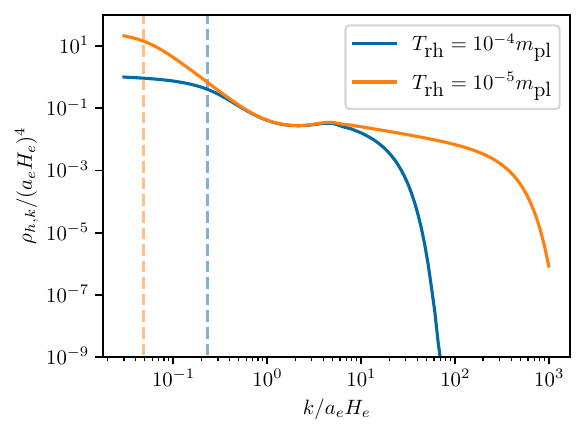}%
    \caption{The phase-space density $\rho_{h,k}$ for $n=2$ and $r=0.01$ for different reheating temperatures $\Trh$. The vertical dashed lines indicate the scale $k = k_{\rm rh} \equiv a_{\rm rh} H(a_{\rm rh})$.}
    \label{fig:TModel-2-Gamma-comp}
\end{figure}

Before presenting the final GW spectrum, it is instructive to first examine the phase-space density $\rho_{h, k} \approx k^4 |\beta_k|^2 / \pi^2$, whose behavior directly reveals the physical origin of the spectral features. This follows directly from Eq.~\eqref{eq:rho_h} upon taking the relativistic limit $\omega_k \to k$. In Fig.~\ref{fig:TModel-2-Gamma-comp}, we show $\rho_{h, k}$ for two reheating temperatures, $\Trh = 10^{-4}\,\mpl$ (blue) and $\Trh = 10^{-5}\,\mpl$ (orange), for the case $n=2$. The vertical dashed lines indicate the scale $k_{\rm rh} \equiv a_{\rm rh} H(a_{\rm rh})$.

Modes with $k \leq k_{\rm rh}$ are not produced during reheating. Instead, they originate from inflation, freeze on superhorizon scales, and re-enter the horizon at or after the end of reheating. Since the GW energy density redshifts as radiation (i.e $\rho_h \propto a^{-4}$), these modes are largely insensitive to the details of reheating. Modes with $k_\text{rh} \leq k \leq k_e$ are also produced during inflation, but re-enters the horizon during reheating. If the  corresponding background equation of state deviates from radiation ($\omega \neq 1/3$), the corresponding energy densities in this case evolve differently, leading to characteristic distortions in the spectrum.

This structure is clearly visible in Fig.~\ref{fig:TModel-2-Gamma-comp}. For $\Trh = 10^{-4}\,\mpl$, the spectrum is approximately flat for $k \leq k_{\rm rh}$ (to the left of the blue dashed line), reflecting the near scale invariance of the inflationary tensor spectrum for modes re-entering during radiation domination. In the intermediate range $k_{\rm rh} < k \leq k_e \equiv a_e H_e$, modes re-enter during reheating when the background behaves as $\omega = 0$, resulting in a suppression of the phase-space density.  For $k \geq k_e$, modes are instead excited during reheating and remain inside the horizon thereafter. In this regime, the spectrum is controlled by particle production from the oscillating background, yielding $|\beta_k|^2 \propto k^{-9/2}$ as shown in Eq.~\eqref{eq:beta_k_app}. This implies $\rho_{h,k} \propto k^{1/2}$ prior to the exponential cutoff.  The same qualitative behavior is observed for a lower reheating temperature, $\Trh = 10^{-5}\,\mpl$, with the corresponding transition scale $k_\text{rh}$ shifted accordingly, as shown by the  orange curve.

\begin{figure}[ht!]
    \centering
    \includegraphics[width=0.45\linewidth]{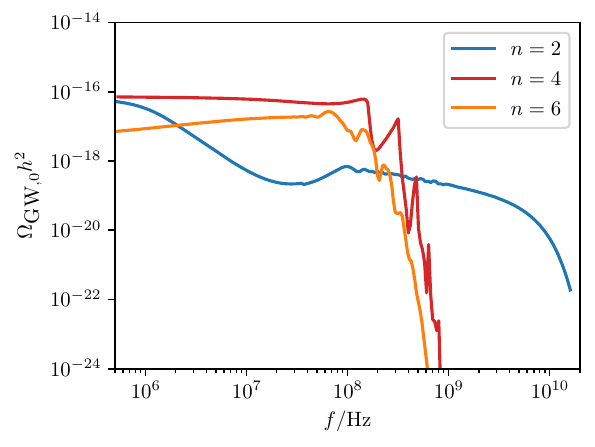}%
    \includegraphics[width=0.45\linewidth]{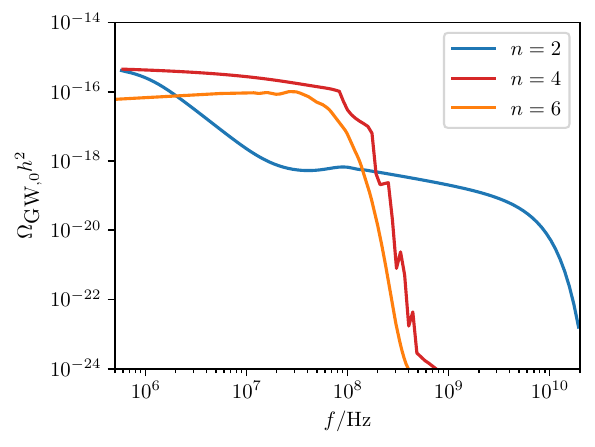}
    \caption{Gravitational-wave spectra for T-model $\alpha$-attractors with different $n$ and $r$. Left: $r=0.001$; right: $r=0.01$. All curves assume $\Trh = 10^{-5}\mpl$.}
    \label{fig:Omega-all-n}
\end{figure}

The resulting GW spectra are shown in Fig.~\ref{fig:Omega-all-n} for two values of the tensor-to-scalar ratio $r$ (left: $r=0.001$, and right: $r=0.01$), with a fixed reheating temperature $\Trh = 10^{-5}\mpl$. 
Since $r \propto H_e^2$, varying $r$ primarily changes the inflationary Hubble scale. The spectral shape, expressed in terms of the dimensionless variable $k/(a_e H_e)$, is largely insensitive to $r$: the T model potential near the origin is independent of $r$ \cite{German:2021tqs}. So the characteristic features appear at similar frequencies, up to a mild shift associated with $f \propto H_e/T_{\rm rh}$ (cf. Eq.~\eqref{eq:f-clean}). By contrast, the overall normalization of the spectrum is set by the inflationary scale, leading to a suppression of the GW amplitude for smaller $r$.

The $n=2$ (the blue curve) case exhibits a qualitatively distinct spectral shape, featuring a dip at intermediate frequencies and a relatively  flat high-frequency tail. This can be traced back to the fact that graviton production is dominated by the oscillatory phase of the inflaton, for which can be described by Boltzmann or Bogoliubov using SPA. The resulting spectrum therefore reflects the coherent oscillatory source, leading to a smoother and less steep high-frequency behavior. We have actually already explained these features in Fig.~\ref{fig:TModel-2-Gamma-comp}.

For steeper potentials ($n=4,6$), the spectra are instead dominated by the non-adiabatic transition near the end of inflation as explained in previous section. In this regime, the high-frequency behavior is governed by the transition contribution, resulting in a steeper fall-off and the absence of the intermediate-frequency dip present in the quadratic case.  We note that these features are consistent with the results shown in Ref.~\cite{Mudrunka:2026kgm}.

The relative normalization between the $n=4$ and $n=6$ spectra follows from the time dependence of the inflaton decay rate, $\Gamma_\phi(t) \propto \rho_\phi^{1/2-1/n}$. For larger $n$, the inflaton energy density redshifts more rapidly, and the decay rate correspondingly decreases faster.  Since both cases start from the same initial energy density and are normalized to the same reheating temperature, reaching this endpoint requires a larger amount of expansion for larger $n$. As a result, the universe remains inflaton-dominated over a longer period. Gravitons produced during this phase are therefore redshifted over a longer interval before reheating completes, leading to a smaller final amplitude. This explains the suppression of the GW signal for $n=6$ relative to $n=4$. 

Finally, all predicted signals lie well below the BBN bound on extra radiation, $\Omega_{\rm GW,0} \lesssim 2 \times 10^{-6}$~\cite{Caprini:2018mtu}. While proposed high-frequency GW detectors probe a similar frequency range, their projected sensitivities remain weaker than the BBN constraint; see e.g. Ref.~\cite{Aggarwal:2025noe}. Although the resulting GW signal is not expected to be observable with such detectors, the structure of the spectrum provides insight into the dynamics of the inflation–reheating transition. In particular, the transition contribution depends on the duration and  the time profile of the non-adiabatic evolution, suggesting that gravitational particle production may serve as a probe of this otherwise inaccessible epoch.
%%%%%%%%%%%%%%%%%%%%%%%%%%%%%%%%%%%%%%%%%%%%%%%%%%%%%%%%%%%%%%%%%%%
\section{Conclusion}\label{sec:conclusion}
%%%%%%%%%%%%%%%%%%%%%%%%%%%%%%%%%%%%%%%%%%%%%%%%%%%%%%%%%%%%%%%%%%%%

Gravitons are inevitably produced through their minimal coupling to a time-dependent cosmological background. After inflation, this production can be sourced by the oscillating inflaton field during reheating. Two complementary frameworks have been widely employed to study this process: the perturbative Boltzmann approach, and the more general Bogoliubov formalism, which describes particle production from an evolving background.

In this work, we have presented a systematic analysis of graviton production during reheating for an inflaton oscillating in a general monomial potential, $V(\phi) \propto \phi^n$ around its minimum. We have compared the Boltzmann and Bogoliubov approaches both numerically and analytically, and clarified their respective regimes of validity. 

The Boltzmann method is based on the phase-space Boltzmann equation for the graviton distribution. The matrix elements are obtained by perturbatively expanding the Einstein--Hilbert action and deriving the interaction vertices, while treating the inflaton as a classical, time-dependent background. Owing to its coherent oscillation, this background can be decomposed into Fourier modes, which act as sources for graviton production. Although the Boltzmann approach ultimately originates from quantum field theory in a time-dependent spacetime, it effectively relies on a local Minkowski approximation and a quasi-static background. Particle production is thus described as perturbative emission from a classical source, with the notion of particles fixed as in flat spacetime. As a result, genuinely non-adiabatic effects associated with the time-dependent background---such as the localized transition at the end of inflation---are not captured. By contrast, the Bogoliubov method directly implements quantum field theory in curved spacetime, where particle production arises from the time dependence of the background. This allows one to capture non-perturbative and non-adiabatic effects, including the transition contribution that is absent in the Boltzmann description.

For a quadratic potential ($n=2$), particle production is dominated by the fast oscillatory phase of the inflaton. In this regime, the Boltzmann and Bogoliubov approaches yield equivalent spectra for short-wavelength modes, as shown in Fig.~\ref{fig:gw-spec-quadratic}. This result is consistent with previous studies~\cite{Kaneta:2022gug, Basso:2022tpd}, which considered scalar particle production and similarly found agreement between the two approaches in the quadratic case.

For steeper potentials with $n>2$, the situation changes qualitatively. The Boltzmann approach fails to reproduce the full spectrum obtained in the Bogoliubov formalism, as illustrated in Figs.~\ref{fig:quartic-spec} and \ref{fig:sextic-spec}. The origin of this discrepancy lies in the non-adiabatic transition from inflation to reheating, which provides the dominant contribution in this regime. This effect is naturally captured by the Bogoliubov formalism but is absent in the Boltzmann treatment. Consequently, the Boltzmann approach does not provide a reliable description of graviton production for steeper inflaton potentials. This distinction is also reflected in the resulting gravitational-wave spectra shown in Fig.~\ref{fig:Omega-all-n}.

To elucidate the physical origin of our numerical results, we have developed analytical approximations within both frameworks. In the Boltzmann approach, we derived a general analytical approximation in Eq.~\eqref{eq:boltzmann-power}, which can reproduce the standard quadratic result in the literature. In the Bogoliubov formalism, we identify two distinct contributions: a fast oscillatory component, described by Eq.~\eqref{eq:beta_k_app_j} using a stationary-phase approximation, and a transition contribution, given by Eq.~\eqref{eq:beta_transition_final}, which captures the localized burst of non-adiabaticity at the end of inflation. These analytical approximations provide a clear guide to the underlying particle production mechanisms. In particular, the transition approximation in Eq.~\eqref{eq:beta_transition_final} offers a simple and robust description of the dominant contribution for $n>2$, and reproduces the corresponding numerical Bogoliubov behavior.

%%%%%%%%%%%%%%%%%%%%%%%%%%%%%%%%%%%%%%%%%%%%%%%%
% This paragraph is for the general reader
%%%%%%%%%%%%%%%%%%%%%%%%%%%%%%%%%%%%%%%%%%%%%%%%
In summary, we have shown that graviton production during reheating is controlled by distinct mechanisms depending on the inflaton potential. While the Boltzmann approach is sufficient for quadratic potentials, steeper potentials require the full Bogoliubov formalism to capture non-adiabatic transition effects. We have derived analytical approximations for both the oscillatory and transition contributions, which reproduce the numerical spectra in the regimes where each mechanism dominates. Although our analysis has focused on gravitons, the underlying physics is general, and the methods developed here can be readily extended to other particle species minimally coupled to a time-dependent cosmological background.
%%%%%%%%%%%%%%%%%%%%%%%%%%
\section*{Acknowledgments}
%%%%%%%%%%%%%%%%%%%%%%%%%%%
We are grateful to Manuel Drees for his  comments on the manuscript as well as helpful discussions. The authors also  thank Nicolás Bernal, Robert Brandenberger, Xiaoyong Chu, Ce Ji, Kunio Kaneta, Marco Peloso, Shi Pi, and Xunjie  Xu for discussions on various aspects of this project. YX was supported by the Natural Sciences and Engineering Research Council (NSERC) of Canada. WBZ was Supported by the National Natural Science Foundation of China (NSFC) under Grant Nos.~12347103, 12547104 and the Fundamental Research Funds for the Central Universities (grant No.\,E5ER6601A2). 

% \clearpage

\appendix

\section{Feynman Rules}\label{sec:feynman-rules}
Here we list the Feynman rules of graviton and scalar-graviton interaction by treating the scalar field as homogeneous and classical. They are adopted from Refs.~\cite{Holstein:2006bh, Choi:1994ax}. $K_j$ and $V_j$ are the Fourier coefficients of the kinetic and potential energy, defined in Eq.~\eqref{eq:K-V-four-decomp}. The Feynman rules are:
\begin{itemize}
    \item Graviton propagator
        \begin{equation}
        \vcenter{\hbox{\includegraphics[scale=0.8]{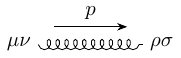}}}
        \sim \frac{i}{2p^2} \left(\eta^{\mu\rho} \eta^{\nu\sigma} + \eta^{\mu\sigma} \eta^{\nu\rho} - \eta^{\mu\nu} \eta^{\sigma \rho} \right).
    \end{equation}
    \item External graviton $\sim \epsilon^{\mu\nu}$
    \item ``Interaction vertex'' of the inflaton condensate and one graviton ($\sim T_\phi^{\mu\nu}$)
    \begin{equation}
        \vcenter{\hbox{\includegraphics[scale=0.7]{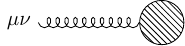}}}
        \sim 
        \frac{-i}{\mpl} \left[ 2\delta_{\mu 0} \delta_{\nu 0} K_j - \eta_{\mu\nu} \left(K_j- V_j \right)  \right].
    \end{equation}
    \item ``Interaction vertex'' of the inflaton condensate and two gravitons 
    \begin{align}
        \vcenter{\hbox{\includegraphics[scale=0.7]{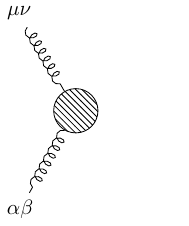}}}
        \sim \frac{4i}{\mpl^2} \Bigg[2I_{\mu \nu 0 \xi} I^{\xi}_{0 \alpha \beta} K_j - \frac{1}{2} \left( \eta_{\mu \nu} I_{00\alpha\beta} + \eta_{\alpha\beta} I_{00\mu\nu} \right) K_j \notag\\
        - \frac{1}{2} \left( I_{\mu\nu \alpha\beta} - \frac{1}{2} \eta_{\mu\nu} \eta_{\alpha\beta}\right) \left(K_j - V_j \right) \Bigg].
    \end{align}
    Actually the first two terms don't contribute, if the gravitons are on-shell (due to the tracelessness). Here we have defined 
    \begin{equation}
        I_{\mu\nu\alpha\beta} = \frac{1}{2} (\eta_{\mu\alpha} \eta_{\nu\beta} + \eta_{\mu\beta} \eta_{\nu\alpha}).
    \end{equation}

    \item Three graviton vertex
    \begin{equation}
\vcenter{\hbox{\includegraphics[width=0.2\linewidth]{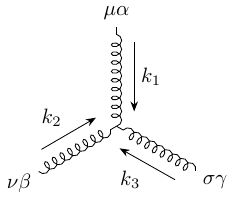}}}
\qquad
\begin{aligned}
        \sim \frac{2i}{\mpl} [\text{Sym}] \biggl[ \frac{1}{2} P_6 (k_1 \cdot k_2 \eta_{\mu\alpha} \eta_{\nu\sigma} \eta_{\beta\gamma}) - P_3 (k_{1\sigma} k_{2\gamma} \eta_{\mu\nu} \eta_{\alpha\beta})  \\
        - 2 P_3(k_1\cdot k_2 \eta_{\alpha \nu} \eta_{\beta\sigma} \eta_{\gamma \mu}) + \frac{1}{2} P_3 (k_1\cdot k_2 \eta_{\mu\nu} \eta_{\alpha \beta} \eta_{\alpha\gamma}) \\
        + P_6(k_{1\sigma} k  _{2\mu} \eta_{\alpha \nu} \eta_{\beta\gamma}) - \frac{1}{4} P_3(k_1 \cdot k_2 \eta_{\mu\alpha} \eta_{\nu\beta} \eta_{\sigma\gamma}) \biggl].
\end{aligned}
\end{equation}
    Here $[\text{Sym}]$ means the symmetrization between $\mu$ and $\alpha$, $\nu$ and $\beta$, and $\sigma$ and $\gamma$ respectively. $P_i$ denotes the $i$ different permutations of the momenta and Lorentz indices \cite{Choi:1994ax}.
    % \item Sum over Fourier modes 
\end{itemize}

We work in the transverse and traceless gauge, so that the graviton polarization tensor follows
\begin{equation}
    k_\mu \epsilon^{\mu\nu} = \eta_{\mu\nu} \epsilon^{\mu\nu} = 0.
\end{equation}
It also satisfies $\epsilon^\sigma_{\mu\nu} \epsilon^{\sigma', \mu\nu} = \delta^{\sigma, \sigma'}$. The graviton polarization sum is given by \cite{deAquino:2011ix,Barman:2023ymn}
\begin{equation}
    \sum_{\sigma} \epsilon^{\sigma, \mu\nu} \epsilon^{*\sigma, \alpha\beta} = 
    \frac{1}{2} \left( \hat\eta^{\mu\alpha}\hat\eta^{\nu\beta} + \hat\eta^{\mu\beta}\hat\eta^{\nu\alpha} - \hat\eta^{\mu\nu} \hat\eta^{\alpha\beta} \right),
    \label{eq:grav_polarisation}
\end{equation}
with
\begin{equation}
    \hat\eta^{\mu\nu} = \eta^{\mu\nu} - \frac{k^\mu \bar k ^\nu + k^\nu \bar k^\mu}{k \cdot \bar k}, \,
    k^\mu = (E_k, \vec{k}), \,
    \bar k^\mu = (E_k, -\vec{k}).
\end{equation}

As shown in Fig.~\ref{fig:graviton-proc}, there are two processes contributing to the graviton production from inflaton condensate. Using the Feynman rules and graviton polarization sum listed above, we find the contributions including the interference term are 
\begin{subequations}
\begin{align}
    \sum_{\sigma, \sigma'}|\mathcal{M}_{1, j}|^2 &= \frac{8\rho_\phi^2}{\mpl^4}| K_j -  V_j|^2\,, \\
    \sum_{\sigma, \sigma'}|\mathcal{M}_{2, j}|^2 &=  \frac{2\rho_\phi^2}{\mpl^4} |2 K_j -  V_j|^2\,, \\
    \sum_{\sigma, \sigma'} 2 \Re{\mathcal{M}_{1, j} \mathcal{M}_{2, j}^*} &= -\frac{8\rho_\phi^2}{\mpl^4} \left(2 |K_j|^2 - 3 K_j V_j^* + |V_j|^2 \right)\,.
    \end{align}
    \label{eq:mat-elements}
\end{subequations}
Summing these terms leads to cancellation of the kinetic term and we end up with the spin-averaged matrix element
\begin{equation}
    \frac{1}{4}\sum_{\sigma, \sigma'} |\mathcal{M}_{\text{tot}, j}|^2 = \frac{\rho_\phi^2(t)|V_j|^2}{2\mpl^4}\,.
\end{equation}

\section{Numerical Details}
\label{app:num}
In this appendix, we present some details on numerical computation, which are omitted in the main text. Depending on the shape of potential around minimum, there is an appropriate choice of time variable for numerical stability \cite{Turner:1983he, Figueroa:2020rrl}. The new time variable $\eta$ is defined as 
\begin{equation}
   \dd{t} =  a^{\tilde{\alpha}} \dd{{{\eta}}} = a \dd{\tau},
\end{equation}
with the $\tilde{\alpha}$ parameter defined as  
\begin{equation}
    \tilde{\alpha} = 3\cdot \frac{n-2}{n+2} = 3 \omega.
\end{equation}
Oscillation periods are approximately constant in the ${\eta}$ time and are roughly the inverse of \cite{Turner:1983he}
\begin{equation}
    \tilde{m}(t=t_e) \sim \sqrt{\lambda} \mpl^{(4-n)/2} \phi_e^{(n-2)/2},
\end{equation}
where the field value at the beginning of oscillation is approximated as the field value at the end of slow roll inflation $\phi_e$. 

The (overcomplete) background equations are now
\begin{align}
\begin{split}
    \partial_\eta^2 \phi + \left( (3-\tilde{\alpha})\mathcal{H} + a^{\tilde{\alpha}} \Gamma_\phi \right)\partial_\eta \phi + a^{2\tilde{\alpha}} \pdv{V(\phi)}{\phi} = 0, \\
    \mathcal{H}_\eta^2 = \left(\frac{\partial_\eta a}{a} \right)^2 = \frac{a^{2\tilde{\alpha}}}{3\mpl^2} \left( \rho_\phi + \rho_r \right), \\
    \partial_\eta \rho_r + 4 \mathcal{H}_\eta \rho_r = a^{\tilde{\alpha}} (1+\omega)\Gamma_\phi \rho_\phi, \\
    \frac{\partial_\eta^2 a}{a} 
    % = \frac{a^{2\alpha}}{6\mpl^2} \left( (2\alpha - 1) \rho_{tot} - 3 p_{tot} \right) 
    = \frac{a^{2\tilde{\alpha}}}{6\mpl^2} \left[ (2\tilde{\alpha} - 1) (\rho_\phi + \rho_r) - 3 \left(p_\phi + \frac{1}{3}\rho_r \right) \right].
\end{split}
\label{eq:background-alpha}
\end{align}
The last equation is superfluous but can be useful to control the accuracy of the solver. Here we defined the new Hubble $\mathcal{H}_\eta = \partial_\eta a /a$ with respect to the new time variable ${\eta}$ and the inflaton energy density and pressure
\begin{equation}
    \rho_\phi = \frac{ \left( \partial_\eta\phi \right)^2 }{2a^{2\tilde{\alpha}}} + V(\phi), \quad p_\phi = \frac{ \left( \partial_\eta\phi \right)^2 }{2a^{2\tilde{\alpha}}} - V(\phi).
\end{equation}
Eqs.~\eqref{eq:background-alpha} are solved numerically to provide the background quantities for further computation of graviton production. Using these equations, we see that the Hubble slow-roll parameter is given by
\begin{equation}
    \epsilon_1 = - \frac{\dot{H}}{H^2} = - \frac{1}{\mathcal{H}^2}\frac{\partial_\eta^2 a}{a} + \tilde{\alpha} + 1,
\end{equation}
which is used to determine the end of inflation. 

Both Boltzmann method (Eq.~\eqref{eq:boltzmann-correction}) and stationary point approximation in the Bogoliubov method (Eq.~\eqref{eq:bogo-ana}) require $\tilde{m}(t)$. With the solution for the background, one can determine the oscillation frequency $\tilde{m}$ by the location of the minima of the field value. Since often the derivatives of $\tilde{m}$ are used, the ``measured'' $\tilde{m}$ often leads large numerical noise. Therefore, a (fourth-order) polynomial fitting of $\ln(\tilde{m})$ as a function of $\ln(a)$ is carried out. This is equivalent of a power law fitting with running$^{3}$. Sub-one-percent relative errors are achieved. The derivatives of $\tilde{m}$ are then computed from the fitted function.

For Boltzmann method, Eq.~\eqref{eq:exact_boltzmann} is implemented by isolating each oscillation and the Fourier decomposition is then computed for each oscillation independently. The crucial quantity in the Bogoliubov calculation is
\begin{equation}
    \frac{a''}{a} = \frac{a^2}{6\mpl^2} \left( 4V - \frac{(\partial_\eta \phi)^2}{a^{2\tilde{\alpha}}} \right).
\end{equation}
The numerical Bogoliubov result is computed via Eqs.~\eqref{eq:EOM-h} and~\eqref{eq:EOM-alpha}, depending on if $a''/a$ becomes negative. If it does, then we find the former more efficient. We also find it useful to control the solver accuracy with the Wronskian condition $h h'^{*} - h^* h' = i$. If Eq.~\eqref{eq:EOM-h} is used, then the exact numerical spectrum can be obtained via Eq.~\eqref{eq:bogo-exact}.

\section{$\alpha$-attractor T model}
\label{app:T_model}
In this work, we adopt the T-model $\alpha$-attractor~\cite{Kallosh:2013hoa, Kallosh:2013maa} as our inflationary setup:
\begin{equation}
    V(\phi) = V_0 \tanh^n \left( \frac{\phi}{\sqrt{6\alpha}\mpl} \right),
\end{equation}
which is characterized by two free parameters, $n$ and $\alpha$. This class of models provides a smooth interpolation between a plateau potential at large field values, relevant for inflation, and an effective monomial potential near the origin, which governs the post-inflationary oscillatory dynamics. The latter property makes it particularly suitable for studying reheating in a controlled manner.

We require the model to be consistent with CMB observations. The parameter $\alpha$ can be determined by inverting the slow-roll expressions for the scalar spectral index $n_s$ and the tensor-to-scalar ratio $r$~\cite{German:2021tqs} 
\begin{equation}
\alpha =
\frac{64}{3}\,
\frac{n^2\,r}{n^2\bigl(8(1-n_s)-r\bigr)^2 - 4r^2}\,.
\end{equation}
which follows from inverting the slow-roll expressions for $n_s$ and $r$ in this model. We adopt the recent ACT result~\cite{AtacamaCosmologyTelescope:2025blo} for the spectral index, $n_s = 0.974$. The current bound on the tensor-to-scalar ratio is $r < 0.036$ at $95\%$ confidence level~\cite{BICEP:2021xfz}, which implies an upper bound $\alpha \lesssim 27$, with only mild dependence on $n$\footnote{The upper bound decreases slightly if one adopts the Planck 2018 best-fit value \cite{Planck:2018vyg}.}. In the limit $r \ll 8(1-n_s)$, the above equation reproduces the universal feature with $\alpha \simeq r/(3(1-n_s)^2)$. Here, the tensor-to-scalar ratio and spectral index can be respectively written as $r\simeq 12 \alpha /N_*^2$ and $1-n_s \simeq 2/N_\star$, where $N_*$ denotes the number of e-folds between the horizon exit at $\phi_*$ (defined below) and end of inflation.

The field value at horizon exit, $\phi_*$, can be obtained from $n_s$ as ~\cite{German:2021tqs}
\begin{equation}
    \cosh^2 \left(\frac{\phi_*}{\sqrt{6\alpha}\mpl}\right) = \frac{1}{2(1-n_s)} \left(1-n_s + \frac{2}{3\alpha} + \sqrt{(1-n_s)^2 + \frac{2(1-n_s)}{3 \alpha} + \frac{4}{9\alpha^2}} \right).
\end{equation}
This relation fixes the position of the field during inflation and determines the slow-roll parameters evaluated at horizon crossing.

Defining $x \equiv \phi/(\sqrt{6\alpha}\mpl)$, the first slow-roll parameter reads
\begin{equation}
    \epsilon_V = \frac{n^2}{12\alpha}\,\sinh^{-2}(x)\cosh^{-2}(x).
\end{equation}
As expected, $\epsilon_V$ is exponentially suppressed at large field values, reflecting the flatness of the potential during inflation, and grows rapidly as the field approaches the origin.

The overall normalization of the scalar power spectrum fixes the scale of the potential
\begin{equation}
    \mathcal{A} = (2.1 \pm 0.1) \times 10^{-9}
    = \frac{1}{24\pi^2} \frac{1}{\epsilon_V} \left.\frac{V}{\mpl^4}\right|_{\phi=\phi_*}.
\end{equation}
This relation depends implicitly on $\alpha$ and $n$ and is used to determine $V_0$. In particular, changing $r$ modifies $V_0$ through its dependence on $\alpha$, thereby shifting the overall energy scale of the model.

To connect with the reheating dynamics, it is useful to expand the potential around $\phi=0$, where the inflaton oscillates after inflation. Using the expansion of $\tanh x$, we obtain
\begin{equation}
    V(\phi) = V_0 \left(\frac{\phi}{\sqrt{6\alpha}\mpl}\right)^n + \mathcal{O}(\phi^{n+2})\,.
\end{equation}
Thus, near the minimum, the potential reduces to an effective monomial form, with the same power $n$ as in the full potential. Matching to Eq.~\eqref{eq:vphi}, we obtain
\begin{equation}
     \lambda = \frac{V_0}{\mpl^4} (6\alpha)^{-n/2}.
     \label{eq:lambda_V0}
\end{equation}
This shows that the effective coupling $\lambda$ is determined by both the normalization $V_0$ and the parameter $\alpha$. In particular, while $V_0$ depends on the inflationary normalization, the combination entering the reheating dynamics involves the rescaled parameter $(6\alpha)^{-n/2}$, reflecting the structure of the potential near the origin.

\section{Comparison between Oscillation and Transition Contributions}
\label{app:dominance}
\begin{figure}[h!]
    \centering
    \includegraphics[width=0.5\linewidth]{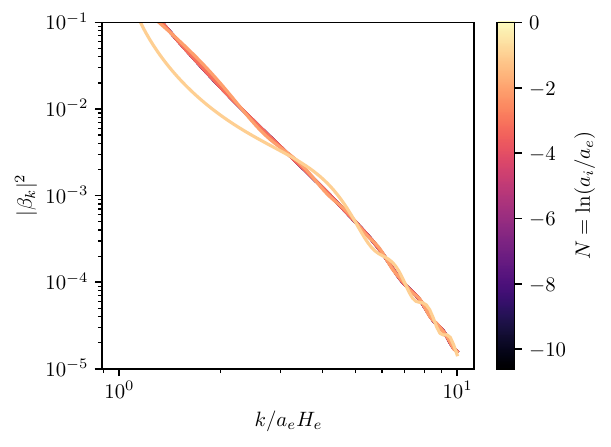}%
    \includegraphics[width=0.5\linewidth]{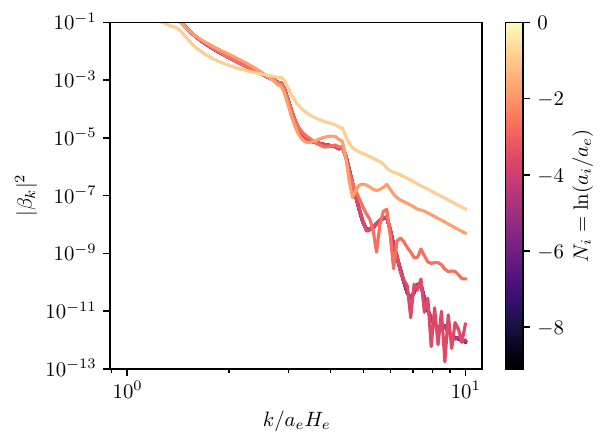}
    \caption{Bogoliubov coefficients $|\beta_k|^2$ with different $\tau_i$'s as the integral lower limit with quadratic potential (left) and quartic potential (right). The lower limit is parametrized as the number of $e$-fold prior to the end of inflation: $N_i = \ln(a(\tau_i)/a_e)$.}
    \label{fig:bogo_fast_all}
\end{figure}
We have argued in the Section \ref{sec:transition}, that the oscillation contribution is suppressed at $n>2$  due to the structure of characteristic momentum and suppression of the Fourier coefficients. On the other hand, the transition contribution is largely unaffected by changing the potential. Furthermore, in Section \ref{sec:transition}, we have derived a simple analytical fit to the transition contribution to the graviton production and we find that $\Delta \tau \sim 0.5 \, a_e H_e$ fits the spectrum pretty well with $n=4, 6$ in Section~\ref{sec:comparison}. 
Here, we want to explicitly demonstrate that transition contribution indeed dominates with higher power potential $n>2$.

First, we start from the approximation for the Bogoliubov coefficients in Eq.~\eqref{eq:bogo-approx}. Since we are mainly interested in the high-$k$ tail of the spectrum with $|\beta_k| \ll 1$, it is sufficient. The integral requires lower and upper limits: by changing the integral limits, we can isolate various contributions and pinpoint precisely the most dominant contribution.

Figs.~\ref{fig:bogo_fast_all} shows the impact of lower integration limit on the final Bogoliubov coefficients. Different lower integration limits are represented as curves with different color: light (dark) corresponds to small (large) $a_i/a_e$. In the regime where $|\beta_k| \ll 1$, Eq.~\eqref{eq:bogo-approx} is equivalent to the Bogoliubov coefficients computed Eq.~\eqref{eq:EOM-h}. And changing the lower limits is equivalent to changing the vacuum state of gravitons: the assumed vacuum state is incorrect, if one starts the integral near the end of inflation, where the non-adiabaticity is more pronounced. However, we see that on the left with $n=2$ potential: even with such incorrect initial vacuum state (lighted curve), the spectrum is still very close to the accurate result (dark curve). This can be easily explained: the oscillation contribution produces the most gravitons and the non-adiabaticity associated with the transition is not consequential in this case. 

On the right, for the $n=4$ potential, the behavior differs qualitatively from the $n=2$ case. Starting the integration closer to $\tau_e$ leads to larger values of $|\beta_k|^2$, as seen from the lighter curves. This can be understood as follows: near $\tau_e$, the transition effects become significant due to the breakdown of adiabaticity. Choosing the initial time closer to $\tau_e$ effectively imposes initial conditions in a more non-adiabatic regime, thereby enhancing the resulting particle production. This behavior reflects the dominance of the transition contribution in this case.

\begin{figure}[h!]
    \centering
    \includegraphics[width=0.5\linewidth]{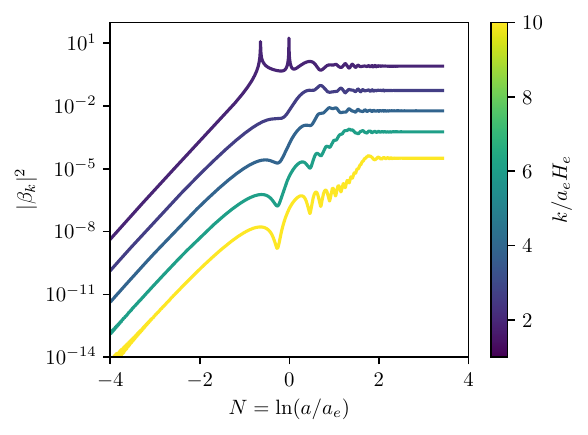}%
    \includegraphics[width=0.5\linewidth]{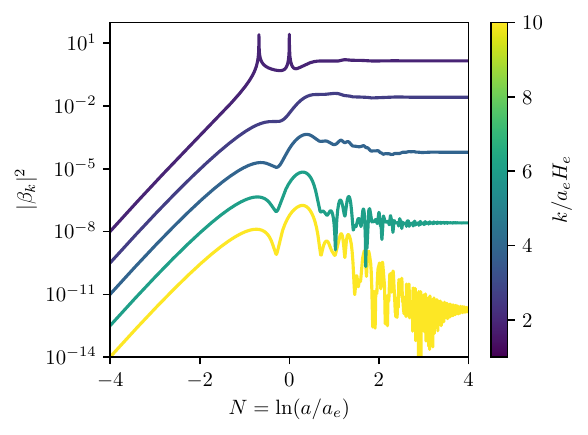}
    \caption{Time evolution of $|\beta_k|^2$ in $n=2$ (left) and $n=4$ potential (right). The color of the curve represents the comoving momentum. In both figures, the momentum goes from $k = a_e H_e$ to $k = 10 \cdot a_e H_e$ from top to bottom. Number of $e$-folds relative to the end of inflation is used for the horizontal axis.}
    \label{fig:beta_time_evo}
\end{figure}

To view the evolution of $\beta_k$ from a complementary perspective, we can also trace its entire time evolution, computed using Eq.~\eqref{eq:EOM-h}, with $n=2$ and $n=4$ potentials. They are shown in Figs.~\ref{fig:beta_time_evo}. Each line corresponds to a single comoving momentum, from $k=a_e H_e$ (topmost) to $k=10 a_e H_e$ (bottommost). The top two lines from both figures have almost identical evolution, as they correspond to low comoving momentum and  insensitive to the oscillation. On the left, one see that even during reheating, the production continues after the end of inflation $N=0$ until the inflaton condensate decays away. Especially for high $k$ (bottom curves), the average $|\beta_k|^2$ increases as it oscillate caused by the background oscillation. On the contrary, we see that the oscillation barely contributes to production in the $n=4$ model. In such case, the Bogoliubov coefficients decreases at late time, as the background moves away from the transition. This confirms our interpretation that oscillation contributes significantly to graviton production with $n=2$ potential, but not with $n=4$ potential.

\bibliographystyle{JHEP}
\bibliography{biblio}
%%%%%%%%%%%%%%%%%%%%%%%%%

\end{document}